\documentclass[numberedappendix, twocolumn]{emulateapj} 
\usepackage{rotating} 
\usepackage{amsmath,amssymb}
\usepackage[usenames,dvipsnames]{pstricks}
\usepackage{epsfig}
\usepackage{subfigure}
\usepackage{pst-grad} 
\usepackage{pst-plot} 
\usepackage{color}

\begin{document}


\title{Scattering polarization in the Ca {\sc ii} Infrared Triplet with Velocity Gradients.}

\author{E.S. Carlin\altaffilmark{1,2}, R.
Manso Sainz\altaffilmark{1,2}, A. Asensio Ramos\altaffilmark{1,2} and J. Trujillo Bueno\altaffilmark{1,2,3}}
\altaffiltext{1}{Instituto de Astrof\'{\i}sica de Canarias, 38205, La Laguna,
Tenerife, Spain}\altaffiltext{2}{Departamento de Astrof\'\i sica, Universidad de La Laguna, Tenerife, Spain}\altaffiltext{3}{Consejo
Superior de Investigaciones Cient\'{\i}ficas, Spain} \email{ecarlin@iac.es}

\begin{abstract}
Magnetic field topology, thermal structure and plasma motions are the three main
factors affecting the polarization signals used to understand our star. In this
theoretical investigation, we focus on the effect that gradients in the macroscopic vertical
velocity field have on the non-magnetic scattering polarization signals, establishing the
basis for general cases. We demonstrate that the solar plasma velocity gradients
have a significant effect on the linear polarization produced by scattering in chromospheric spectral lines.
 In particular, we show the impact of
velocity gradients on the anisotropy of the
radiation field and on the ensuing fractional alignment of the Ca {\sc ii} levels,
and how they can
lead to an enhancement of the zero-field linear polarization signals. This
investigation remarks the importance of knowing the
dynamical state of the solar atmosphere in order to correctly interpret
spectropolarimetric measurements, which is important, among other things, for
establishing a suitable zero field reference case to infer magnetic fields
via the Hanle effect.
\end{abstract}


\section{Introduction} 
Over the last few years it has become increasingly clear that
the determination of the magnetic field in the ``quiet" solar chromosphere
requires measuring and interpreting the linear polarization profiles
produced by scattering in strong spectral lines, such as H$_{\alpha}$ and
the 8542 \AA\ line of the infrared (IR) triplet of Ca {\sc ii} (e.g., see reviews by
Trujillo Bueno 2010
and Uitenbroek 2011). In these chromospheric lines, the maximum fractional linear polarization signal occurs
at the center of the spectral line under consideration, where the Hanle effect 
(i.e., the magnetic-field-induced modification of the scattering line polarization) operates (Stenflo 1998). Since
the opacity at the center of such chromospheric lines is very significant, it is
natural to find that the response function
of the emergent scattering polarization to magnetic field perturbations peaks in
the upper chromosphere (\v{S}t\v{e}p\'{a}n \& Trujillo Bueno).
This contrasts with the circular polarization signal caused by the Zeeman effect
whose response function
peaks at significantly lower atmospheric heights (Uitenbroek \& Socas Navarro 2004). Of particular importance for
developing the Hanle effect as a
diagnostic tool of chromospheric magnetism is to understand and calculate
reliably the linear polarization profiles
corresponding to the zero-field reference case.

The physical origin of the scattering line polarization is atomic level
polarization
(that is, population imbalances and/or coherence between the magnetic sublevels
of a degenerate level with total angular momentum $J$). Atomic polarization, in turn, is induced by anisotropic radiation pumping, which can be
particularly efficient in the low-density regions of stellar atmospheres where
the depolarizing role of
elastic collisions tends to be negligible. The larger the anisotropy of the incident field,
 the larger the induced atomic level polarization, and the larger the amplitude of the 
 emergent linear polarization.

The degree of anisotropy of the spectral line radiation within the solar
atmosphere
depends on the center-to-limb variation (CLV) of the incident intensity. In a
static model atmosphere
the CLV of the incident intensity is established by the gradient of the source
function of the spectral line under consideration \citep{jtb-sacpeak,ll04}.
However, stellar chromospheres are highly dynamic systems, with shocks and wave
motions (e.g., Carlsson \& Stein 1997). The
ensuing macroscopic velocity gradients and Doppler shifts might have a
significant
impact on the radiation field anisotropy and, consequently, on the emergent
polarization profiles. Therefore, it is important to investigate
the extent to which macroscopic velocity gradients may modify the amplitude and
shape of the emergent linear polarization profiles produced by
optically pumped atoms in the solar atmosphere. The main aim of this first paper
is to explain why atmospheric velocity gradients may
modify the anisotropy of the spectral line radiation and, therefore, the
emergent scattering line polarization. We aim also at evaluating, with the help
of
ad-hoc velocity fields introduced in a semi-empirical solar model atmosphere,
their possible impact
on the scattering polarization of the IR triplet of Ca {\sc ii}. A recent
investigation by Manso Sainz \& Trujillo Bueno (2010), based on radiative
transfer calculations in static model atmospheres, shows why the differential
Hanle effect in these lines is of great potential interest for the exploration
of chromospheric magnetism.


\section{Formulation of the problem and relevant equations}\label{sec:problem} 

\subsection{The atomic model and the statistical equilibrium equations (SEE)}

We assume an atomic model consisting of the five lowest energetic, fine
structure levels of Ca {\sc ii} (see Figure 1). 
The excitation state of the atomic system is given by the populations of its 18
magnetic sublevels and the coherences among them.
We neglect coherences between different energy levels of the same term
(multilevel approximation). 
Moreover, since the problem we consider here (plane-parallel, non-magnetic
atmosphere with vertical velocity fields) is axially symmetric around the
vertical direction, no coherences between different magnetic sublevels exist when
the symmetry axis is taken for quantizing the angular momentum. 
We use the multipolar components of each $J$-level, 
\begin{equation}\label{eq01}
\rho^K_0=\sum_{M=-J}^{+J}(-1)^{J-M}\sqrt{2K+1}\left(
\begin{array}{ccc}
J & J & K \\
M & -M & 0
\end{array}
\right)N_{M},
\end{equation}
where $K=0, ..., 2J$, $N_M$ is the population of the sublevel with magnetic quantum number $M$, and 
the symbol between brackets is the Wigner $3j$-symbol (e.g., Brink \& Satchler 1968). 
Due to the symmetry of the scattering process (no magnetic field, no polarized
incident radiation in the atmosphere's boundaries), in a given level
$N_{+M}=N_{-M}$, 
and the excitation state of the system is described by just 9 independent sublevel
populations.
Consequently, odd-$K$ elements (orientation components) in Equation 1 vanish for
all five levels, and the only independent variables of the problem in the
spherical components formalism are the total populations of the five levels
($\sqrt{2J+1}\rho^0_0$); the alignment components ($\rho^2_0$) of levels 2, 3,
and 5; and $\rho^4_0$ of level 3, whose role is negligible for
our problem.

The statistical equilibrium equations accounting for the radiative and
collisional excitations and deexcitations in the 5-level system of Figure 1 are
given explicitly in Manso Sainz \& Trujillo Bueno (2010 ;hereafter MSTB2010). We have particularized them to the no-coherence case (only $\rho^K_0$
elements) in Appendix \ref{app:B}.
The statistical equilibrium equations for the $\rho^0_0$ components contain terms 
that are equal to those appearing in the statistical equilibrium equations for the populations in a
standard (no polarization) NLTE problem (e.g., Mihalas 1978), plus higher order
terms $\sim J^2_0\rho^2_0$ (see Eqs.~\ref{see01}-\ref{see05}).
The statistical equilibrium equations for the alignment ($\rho^2_0$ components)
are formally similar to the ones for the populations with additional terms $\sim
J^2_0\rho^0_0$ accounting for the generation of alignment from the anisotropy of
the radiation field, and (negligible) higher order terms $\sim J^2_0\rho^2_0$
and $J^2_0\rho^4_0$ (see Equations~(\ref{see06}), (\ref{see07}) and
(\ref{see08})). These equations are expressed in the atom reference frame
(comoving system).


\begin{figure}
\begin{center}
\plotone{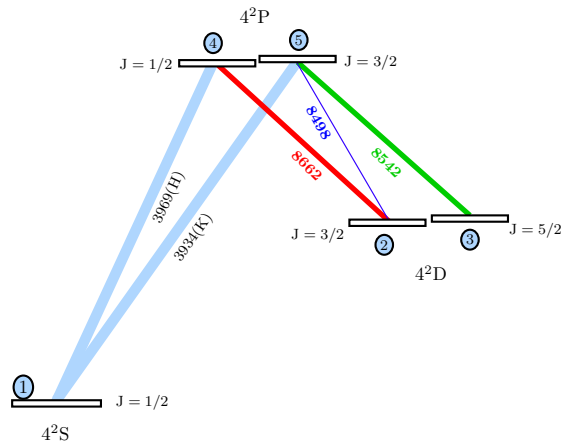}
\end{center}
\caption{Atomic model with energy levels for Ca{\sc ii}. The labels indicate the spectroscopic terms. 
Numbers inside blue filled circles identificate each level. Solid lines connecting 
levels show the allowed radiative transitions and the numbers at the middle of each 
segment are their wavelength in \AA. The wider the width of each connecting line the larger the spontaneous emission rate $A_{ul}$ of the transition. Atomic data for each spectral 
line are shown in Table \ref{tab:atomic}.\label{fig:elevels}}
\end{figure}

 \begin{table}
 \label{tab:atomic}
\centering
\begin{tabular}{cccccc}
\hline
 $\lambda$ (\AA\ ) & $u$ &
${\ell}$ & $A_{u{\ell}}$
(s$^{-1}$) & $ w_{J_{\ell}J_u}^{(2)}$ & $
w_{J_uJ_{\ell}}^{(2)}$ \\
\hline
\multicolumn{6}{c}{Allowed transitions} \\
\hline
3934 (K)    & 5 & 1 & $1.4\times 10^8$  & $0 $ & $\sqrt{2}/2$\\
3969 (H)    & 4 & 1 & $1.4\times 10^8$  & $0$ & $0$\\
8498 & 5 & 2 & $1.11\times 10^6$ & $-2\sqrt{2}/5$ & $-2\sqrt{2}/5$\\
8542 & 5 & 3 & $9.6\times 10^6$  & $\sqrt{7}/5$ & $\sqrt{2}/10$\\
8662 & 4 & 2 & $1.06\times 10^7$ & $\sqrt{2}/2$ & $0$\\
\hline
\end{tabular}

\caption{Short list of Ca {\sc ii} atomic data parameters. From left to right: the central wavelength, the upper and the lower level of each transition, the radiative rates from NIST atomic spectra database ({\tt http://www.nist.gov/physlab/data/asd.cfm}) and the atomic polarizability coefficients introduced by Landi Degl'Innocenti (1984).}
\end{table}

Since the radiation field is axially symmetric, just two radiation field tensor
elements ($J^0_0$ and $J^2_0$) are necessary to describe the symmetry properties of
the spectral line radiation. 
Let $I(\nu, \mu)$ and $Q(\nu, \mu)$ be the Stokes parameters expressed in the observer's frame
at a given height $z$, where $\nu$ is the frequency, $\mu=\cos\theta$ and $\theta$ is the angle that the ray forms with the
vertical direction. Then, the corresponding values seen by a comoving
frame with vertical velocity $v_z$ with respect to the observer's frame are $I'(\nu',
\mu)=I(\nu, \mu)$ and $Q'(\nu', \mu)=Q(\nu, \mu)$, where
$\nu'=\nu(1-v_z\mu/c)$ and $\nu=\nu'(1+v_z\mu/c)$ (to first order in $v_z/c$).
Therefore, the mean intensity at the considered height, can be expressed from
one or another reference frame as:
\begin{align}\label{eq02}
\bar{J}^0_0&=\frac{1}{2}\int d\nu \int_{-1}^1 d\mu \phi'_{\ell u}(\nu, \mu)
I(\nu, \mu) \notag \\
&=\frac{1}{2}\int d\nu' \int_{-1}^1 d\mu \phi_{\ell u}(\nu')
I(\nu'(1+v_z\mu/c), \mu),
\end{align}
where $\phi_{\ell u}(\nu)$ is the absorption profile (e.g., for a Gaussian profile, we would have $\phi_{\ell u}(\nu)=\pi^{-1/2}{\Delta\nu_D}^{-1}\exp(-(\nu-\nu_0)^2/{\Delta\nu_D}^2)$, with
$\nu_0$ the central line frequency and $\Delta\nu_D$ the Doppler width)  and
$\phi'_{\ell u}(\nu, \mu)=\phi_{\ell u}(\nu(1-v_z\mu/c))$, with $v_z>0$ for
upflowing material.
Analogously, the anisotropy in the observer's frame:
\begin{eqnarray}\label{eq03}
\bar{J}^2_0&=&\frac{1}{4\sqrt{2}}\int d\nu \int_{-1}^1 d\mu \phi_{\ell u}'(\mu,
\nu) \nonumber \\
&\times& [(3\mu^2-1)I(\nu, \mu)+3(1-\mu^2)Q(\nu, \mu)],
\end{eqnarray}

The important quantity that controls the ability of an anisotropic radiation
field to generate atomic polarization is the line anisotropy factor for each transition, which
can be calculated as:
\begin{equation}\label{eq:aniso}
{\sc}w_{\mathrm{line}}={\sqrt{2}} \frac{\bar{J}^2_0}{\bar{J}^0_0}.
\end{equation}
Its range goes from $\sc w_{\mathrm{line}}=-0.5$ (for a radiation field coming
entirely from the horizontal plane) to $\sc w_{\mathrm{line}}=1$ (for a collimated vertical
beam). 
\\
\subsection{The radiative transfer equations (RTE)}\label{subsec:rte}

Due to symmetry, in a non-magnetized plane-parallel medium with a vertical
velocity field, light can only be linearly polarized parallel or perpendicularly
to the stellar limb. 
Therefore, chosing the reference direction for positive $Q$ parallel to the
limb, the only non-vanishing Stokes parameters are $I$ and $Q$, and they satisfy
the following radiative transfer equations:

\begin{subequations}\label{eq:rte}
\begin{align}
\frac{\rm d}{{\rm d}s}I&=\epsilon_I-\eta_II-\eta_QQ, \label{rte1}
\displaybreak[0] \\
\frac{\rm d}{{\rm d}s}Q&=\epsilon_Q-\eta_QI-\eta_IQ, \label{rte2}
\end{align}
\end{subequations}

where $s$ is the distance along the ray. The absorption and emission coefficients
are (MSTB2010):
\begin{subequations}\label{eq:coefs}
\begin{align}
\epsilon_I &= \,\epsilon_I^{\rm cont}+\epsilon_I^{\rm line}  \nonumber \\
&=\, {\eta_{I}}^{\rm
cont}B_{\nu}+ \epsilon_0 \left[\rho^0_0(u)+w_{J_uJ_{\ell}}^{(2)}
\frac{1}{2\sqrt{2}} (3\mu^2-1)\rho^2_0(u)\right], \label{coefs1}
\displaybreak[0] \\
\eta_I &= \,\eta_I^{\rm cont}+\eta_I^{\rm line}  \nonumber \\
&=\,{\eta_{I}}^{\rm cont}+\eta_0
\left[\rho^0_0(\ell)+w_{J_{\ell}J_u}^{(2)} \frac{1}{2\sqrt{2}}
(3\mu^2-1)\rho^2_0(\ell)\right], \label{coefs2}
\end{align}
\end{subequations}
\begin{subequations}\label{eq:coefs2}
\begin{align}
\epsilon_Q\,=\,\epsilon^{\rm line}_Q\,=\,\epsilon_0 w_{J_uJ_{\ell}}^{(2)}
\frac{3}{2\sqrt{2}}(1-\mu^2)\rho^2_0(u), \label{coefs3}
\displaybreak[0] \\
\eta_Q\,=\,\eta^{\rm line}_Q\,=\,\eta_0 w_{J_{\ell}J_u}^{(2)}
\frac{3}{2\sqrt{2}}(1-\mu^2)\rho^2_0(\ell), \label{coefs4}
\end{align}
\end{subequations}
where ${\eta_{I}}^{\rm cont}$ and ${\epsilon_{I}}^{\rm cont}$ are the continuum
absortion and emission coefficients for intensity, respectively. Likewise, $\eta_I^\mathrm{line}$ and $\eta_Q^\mathrm{line}$
are the line absorption coefficients for Stokes $I$ and $Q$, respectively, while $\epsilon_I^\mathrm{line}$ and 
$\epsilon_Q^\mathrm{line}$ are the line emission coefficients for Stokes $I$ and $Q$, respectively. The coefficients $w_{J_{\ell}J_u}^{(2)}$ and
$w_{J_{u}J_{\ell}}^{(2)}$ depend only on the transition and are detailed in
Table (\ref{tab:atomic}).The subscripts $u$ and $\ell$ refer to the upper and
lower level of the transition considered, respectively, and $B_{\nu}$ is the
Planck function at the central frequency  $\nu_0$ of the transition. Note also
that:
\begin{subequations}\label{eq:epseta}
\begin{align}
\epsilon_0=\frac{h\nu}{4\pi}A_{u{\ell}}{\phi^{\prime}_{\ell u}(\mu,\nu)}{\cal
N}\sqrt{2J_u+1}, \label{epseta1}
\displaybreak[0] \\
\eta_0=\frac{h\nu}{4\pi}B_{{\ell}u}{\phi^{\prime}_{\ell u}(\mu,\nu)}{\cal
N}\sqrt{2J_{\ell}+1}, \label{epseta2}
\end{align}
\end{subequations}
 where $\cal N$ is the total number of atoms per unit volume. 
 
With the total absortion coefficient for the intensity, the line of sight (los)
optical depth for each frequency is calculated by the following integral along
the ray:
\begin{equation}
\tau^{\rm los}_{\nu}=- \int \eta_I(\mu^{\rm los},\nu) \frac{dz}{\mu^{\rm los}}
\end{equation}

\subsection{Numerical method}
The solution to the non-LTE problem of the second kind considered here (the self-consistent solution of
the statistical equilibrium equations for the density matrix elements together with the
radiative transfer equations for the Stokes parameters) is carried out by 
generalizing the computer program developed by \cite{manso-trujillo-03}, 
to allow for radial macroscopic velocity fields. For integrating the RTE, a
parabolic short-characteristics scheme \citep{kunasz-auer} is used. At each
iterative step, the radiative transfer equation is solved, and $\bar{J}^0_0$ and
$\bar{J}^2_0$ are computed and used to solve the SEE following the accelarated
Lambda iteration method outlined in the Appendix of MSTB2010. Once the
solution for the multipolar components of the density matrix is consistently
reached, the emergent Stokes parameters are calculated for the desired line of
sight, which in all the figures of this paper is $\mu=0.1$.   
This final step is done increasing the frequency grid resolution to a large value
in order to correctly sample the small features and peaks of the emergent profiles.

Some technical considerations have to be kept in mind for the treatment of
velocity fields. Due to the presence of Doppler shifts, the wavelength axis used
to compute $\bar{J}^0_0$ and $\bar{J}^2_0$ must include the required extension
and resolution, because the spectral line radiation may be now shifted and
asymmetric. In our strategy for the wavelength grid, the resolution is larger in
the core than in the wings, keeping the same wavelength grid for all heights.

The cutoff wavelength for the core (where resolution is appreciably higher) is
dictated by the maximum expected Doppler shift. Thus, the core bandwidth is
estimated allowing for a range of $2V_{\rm{max}}$ around the zero
velocity central wavelength of the lines, with
$V_{\rm{max}}$ the maximum velocity found in the atmosphere (in Doppler units).
Apart from that, a minimum typical resolution for the core is set to 2 points
per Doppler width ($\Delta\nu_{D}$). Then, the height with the smallest Doppler
width determines the core resolution, and the height with the maximum
macro-velocity states its bandwidth. 

Furthermore, as frequencies and angles are inextricably entangled (through terms
$\nu-v_z\mu\nu/c$ appearing in the absortion/emission profiles due to the
Doppler effect, like in Eq. (\ref{eq02})), the maximum angular increment ($\Delta\mu_{\rm{max}}$) is
restricted by the maximum frequency increment ($\Delta x_{\rm{max}}\thickapprox
1/2$, in Doppler units). Thus, it must occur that $\Delta\mu_{\rm{max}} \cdot
V_{\rm{max}}\leq1/2$. In the worst case, the maximum allowed angular increment
will be smaller (more angular resolution needed) when the maximum vertical
velocity increases. Besides this consideration, the maximum angular increment
could be even more demanding because of the high sensitivity of the polarization
profiles to the angular discretization.

Finally, the depth grid must be fine enough, in such a way that the maximum
difference in velocity between consecutive points is not too large, the
typical difference being equal to half the Doppler width ($|V(z_i)-V(z_{i-1})|\leq1/2$). If the
difference is larger, the absortion/emission profiles would change abruptly with height, producing
imprecisions in the optical depth increments (Mihalas, 1978).

\section{Effect of a velocity gradient on the radiation field anisotropy and the
mean intensity}\label{sec:preview} 

As we shall see, the presence of a vertical velocity gradient in an atmosphere
enhances the anisotropy of the radiation field and, hence, of the scattering
line polarization patterns. The fundamental process underlying this mechanism can be simply understood with
the following basic examples.

\subsection{Anisotropy seen by a moving scatterer.}\label{sub:scattercloud}
\begin{figure}
\plotone{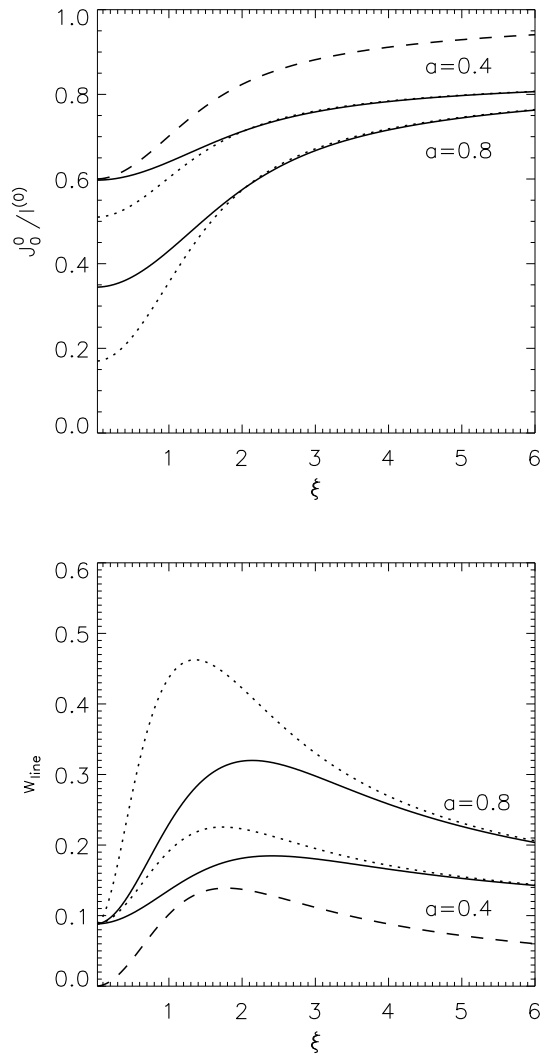}
\caption{$\bar{J}^0_0/I^{(0)}$ (upper panel) and $\sqrt{2}\bar{J}^2_0/\bar{J}^0_0$ (lower panel) 
as a function of the adimensional velocity $\xi$ calculated using an incident line profile 
as in Eq.~(\ref{eq04}) with $u=0.3$ and $a=0.4$ or 0.8 (see labels). 
Dotted lines have been computed for the case of an infinitely sharp absorption profile ($\alpha=0$).
Solid lines refer to the case $\alpha=0.9$ (non saturated line).
The case with no limb darkening ($u=0$) and $a=0.4$ has been plotted for comparison (dashed lines).\label{fig:basic}}
\end{figure}

Consider an absorption spectral line with a Gaussian profile emerging from a
\textit{static} atmosphere with a linear limb darkening law,
\begin{equation}
I(\nu, \mu)=I^{(0)}(1-u+u\mu)[1-a\exp(-(\frac{\nu-\nu_0}{w})^2)], \label{eq04}
\end{equation}
where $I^{(0)}$ is the continuum intensity at disk center, $u$ is the limb
darkening coefficient, $a<1$ measures the intensity depression of the line, and
$w$ its width. In this approximation we assume that all the parameters are constant. 
Now imagine that, at the top of the atmosphere, there is a thin cloud scattering
the incident light given by Eq. (\ref{eq04}) and moving radially at velocity
$v_z$ with respect to the bottom layers of the atmosphere, supposed static. 
We will assume that the absorption profile is Gaussian (dominated by Doppler
broadening), with width $\Delta\nu_D$. When $\Delta\nu_D\ll w$, the incident spectral
line radiation is much broader than the absorption profile (in fact, for
$\alpha=\Delta\nu_D/w=0$ the absorption profile is formally a Dirac-$\delta$
function).
Then, from Equations~(\ref{eq02})-(\ref{eq03}) we can derive explicit
expressions for the mean intensity and anisotropy of the radiation field as a
function of the scatterer velocity (see Appendix A): $\bar{J}^0_0={\cal
I}_0(\alpha;\xi)/2$, $\bar{J}^2_0={\cal I}_2(\alpha;\xi)/4\sqrt{2}$, where the
${\cal I}_{0, 2}$ functions are defined by
Equations~(\ref{eqa04})-(\ref{eqa05}).
The behaviour of $\bar{J}^0_0$ and $\bar{J}^2_0$ with the adimensional velocity
$\xi=v_z\nu_0/(cw)$ (Figure \ref{fig:basic}), is most clearly illustrated in
their asymptotic limits at low velocities:
\begin{align}
\bar{J}^0_0&=\frac{1}{4}(1-\frac{a}{\sqrt{1+\alpha^2}})(2-u) \nonumber \\
+&\frac{a(4-u)}{
12(1+\alpha^2)^{3/2}}\xi^2 + {\rm O}(\xi^3), \label{eq05} \\
\sqrt{2}\frac{\bar{J}^2_0}{\bar{J}^0_0}&=\frac{u}{4(2-u)} \nonumber \\
+&\frac{a(64-56u+7u^2)}{
120(2-u)^2(1+\alpha^2)(\sqrt{1+\alpha^2}-a)}\xi^2  + {\rm O}(\xi^3).
 \label{eq06}
\end{align}
Equation (\ref{eq05}) shows that, for an absorption line ($a>0$), $\bar{J}^0_0$
is always increasing with the velocity since the coefficient of $\xi^2$ is
positive, and this, regardless of the sign of $v_z$ (i.e., regardless of whether the scatterers move upwards
or downwards); if the line is in emission ($a<0$), $\bar{J}^0_0$ monotonically
decreases.
These are the {\em Doppler brightening} and {\em Doppler dimming} effects
\citep[e.g.,][]{ll04}.
An analogous analysis applies to $\bar{J}^2_0$ (Eq.~\ref{eq06}).
Note that, in the absence of limb darkening ($u=0$), the anisotropy vanishes in
a static atmosphere, while the mere presence of a relative velocity between the
scatterers and the underlying static atmosphere induces anisotropy in the radiation
field ---hence, a polarization signal.  
A real atmosphere could then be understood as a superposition of scatterers 
that modify the anisotropy depending on the local
velocity gradient and the illumination received from lower shells.
An interesting discussion on the effect of velocities with directions other than
radial can be found in Landi Degl'Innocenti \& Landolfi (2004; Section 12.4).

\begin{figure}[!b]
\plotone{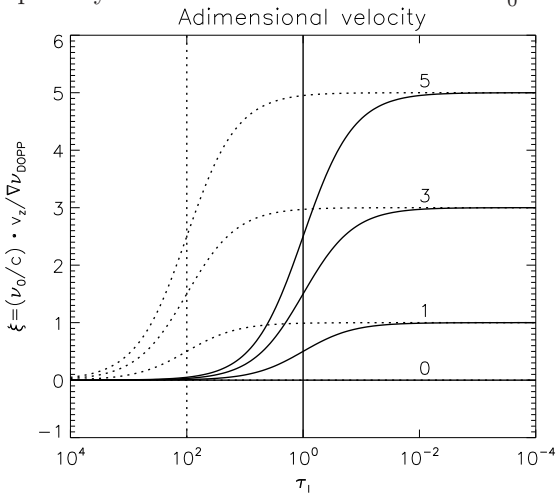}
\caption{Adimensional velocity fields considered in this section. They are
parameterized by the limiting value at small optical depths (labels) and by the location
in optical depth of the largest velocity gradient region (vertical lines marking $\tau_0$). Solid lines: $\tau_0=1$. Dotted lines: $\tau_0=100$. \label{fig:velocity}}
\end{figure}

\begin{figure}[!t]
\plotone{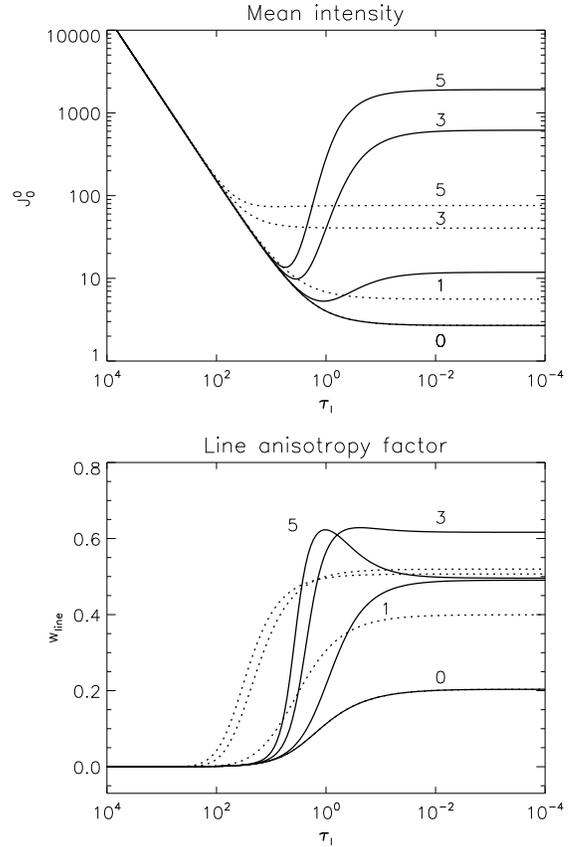}
\caption{$\bar{J}^0_0$ (upper panel) and $\sqrt{2}\bar{J}^2_0/\bar{J}^0_0$ (lower panel) as a 
function of the integrated static line opacity ($\tau_l$) in an expanding atmosphere with 
$S=S^{(0)}(1+\beta\tau_l)$ and different velocity stratifications $\xi=\xi_0/(1+\tau_l/\tau_0)$ (see Fig. \ref{fig:velocity}).
The parameters used in this plot are $\beta=3/2$ and $\kappa_c/\kappa_l=10^{-4}$. The labels indicate the value of $\xi_0$.
Solid lines correspond to $\tau_0=1$ while dotted lines refer to $\tau_0=100$.
\label{fig:me_fig}}
\end{figure}

\begin{figure}[!t]
\plotone{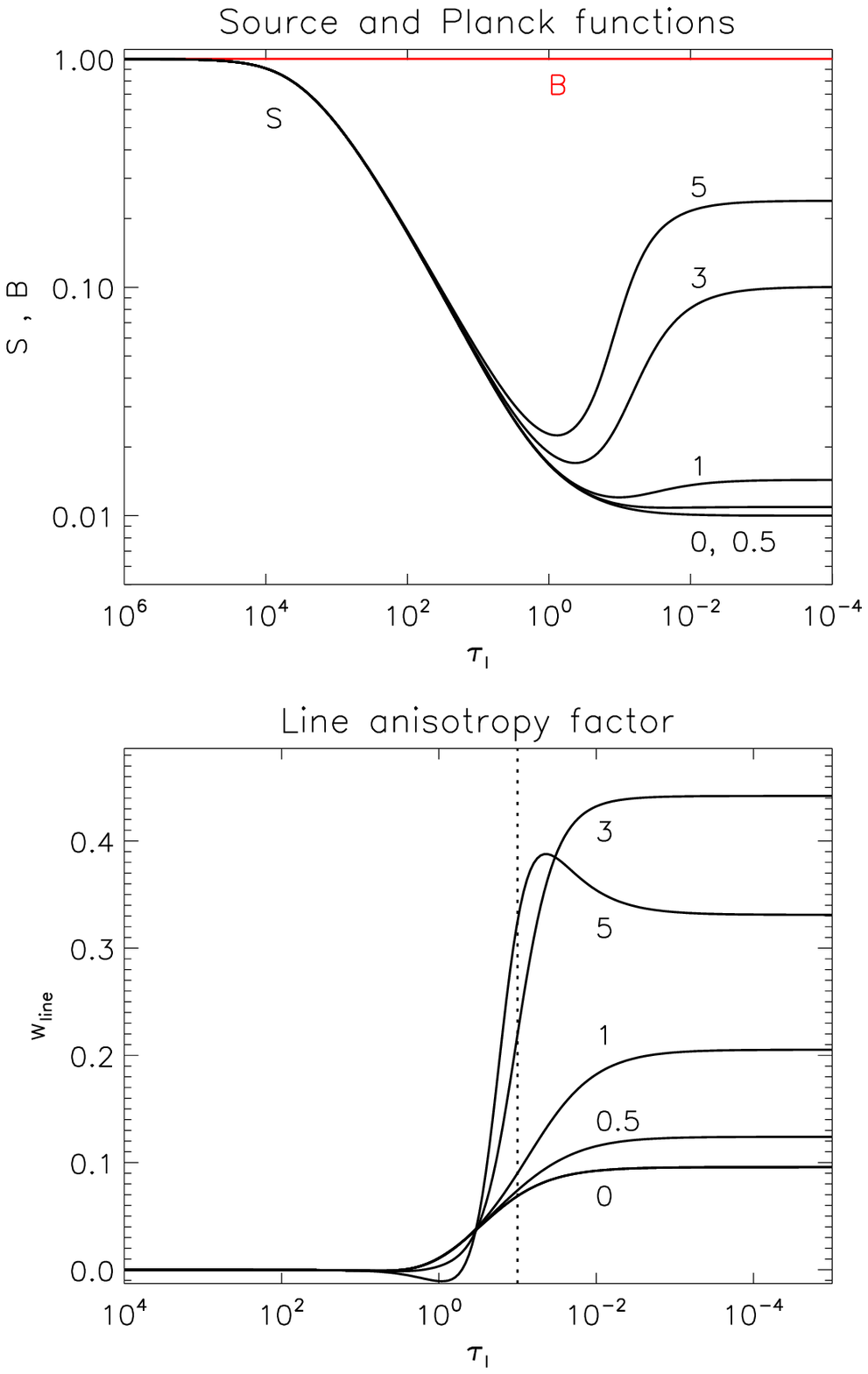}
\caption{Same as Fig. \ref{fig:me_fig} for an isothermal moving 
atmosphere with a two-level atom model using the velocity fields of Fig. \ref{fig:velocity}. We assume a very strong line ($r_c=0$)  
and $\epsilon=10^{-4}$. Upper panel: the Doppler brightening effect increases
the surface source function value. In a two-level model atom in a static atmosphere (label 0) 
this value follows the well known expression $S(0)=\sqrt{\epsilon}B$. The vertical 
axis is in units of the Planck function. Lower panel: we show the amplification 
of the radiation field anisotropy when the velocity gradient increases. The vertical 
line marks the location of the largest velocity gradient ($\tau_l=\tau_0=0.1$). The curve
labeled with ``5'' shows the saturation of the anisotropy and the bump produced by the strong 
velocity gradient taking place at optically thin regions.\label{fig:tla1}}
\end{figure}

Clearly, all the above discussion depends on the Doppler shift induced by the
velocity $v_z$ normalized to the width of the spectral line, i.e., on $\xi$.
A large velocity gradient on a broad line has the same effect as that of a
smaller velocity gradient on a narrow line. 
This is important to be kept in mind since the response of different spectral
lines to the same velocity gradient will be different, what can help us to
decipher the velocity stratification. 
Even different spectral lines belonging to the same atomic species may have very
different widths, as for example, the Ca~{\sc ii} IR triplet and the UV doublet studied in the next
section.

\subsection{Calculations in a Milne-Eddington model.}
The discussion above explains the basic mechanism by means of which a velocity gradient
enhances the anisotropy of the radiation field.
Now we can get further insight on the structure of the radiation field within an
atmosphere with velocity gradients from just the formal solution of the RT
equation for the intensity (e.g., Mihalas 1978).
As before, we neglect effects due to polarization and $J^0_0$ and $J^2_0$ are
calculated from Stokes $I$ alone.
We consider a semi-infinite, plane-parallel atmosphere with a source function
$S=S_0(1+\beta\tau_l)$, where $\tau_l$ is the integrated line optical depth in
the static limit (hence, the element of optical depth
$d\tau_\nu=(r+\phi[\nu(1-v_z(\tau_l)\mu/c)])d\tau_l$, where
$r=\kappa_c/\kappa_l$ is the ratio of continuum to line opacity). 
We begin by considering a vertical velocity field
$v_z(\tau_l)=v_0/[1+(\tau_l/\tau_0)]$ (positive outward the star), shown
in Fig. \ref{fig:velocity}.
Equivalently, we may express the velocity in adimensional terms by using 
$\xi=(\nu_0/c)v_z/\Delta\nu_D$ (the width $\Delta\nu_D$ of the Gaussian
absorption profile is assumed to be constant with depth). The parameter $\tau_0$
fixes the position of the maximum velocity gradient. 
Note that the wavelength dependence of the Doppler effect ($\Delta\lambda_z =
\lambda_0 v_z / c$) is cancelled in the adimensional problem, where velocities
are measured in Doppler units.
It is easy to calculate numerically $I_z(\nu, \mu)$ at every point in the atmosphere and
thus, the mean intensity and anisotropy of the radiation field (Figure
\ref{fig:me_fig}). 

The rise in $\bar{J}^0_0$ in higher layers with respect to the static case
corresponds to the Doppler brightening discussed above. Thanks to the Doppler
shifts, the atoms \textit{see} more and more of the brighter continuum below, which
enhances $\bar{J}^0_0$. When the maximum velocity gradient takes place at 
optically thick enough layers ($\tau_0 \gtrsim 1$), $\bar{J^0_0}$ is also larger than 
for the static case, but it decreases monotonically with height in the atmosphere ($\tau_0 = 10^2$,
dotted lines in upper panel of Fig. \ref{fig:me_fig}). Note that the important quantity that
modulates the increase in $\bar{J^0_0}$ is not the maximum velocity
but the velocity gradient (difference in velocity between optically thick and optically thin parts of the atmosphere). The larger the gradient, the
more pronounced the radiative decoupling is between different heights. An extreme
example of such radiative decoupling could be found in supernovae explosions, where the
vertical velocity gradients are huge.

In our case (vertical motions), the Doppler brightening implies an
enhancement of the contribution of vertical radiation to Eq. (\ref{eq03}) with
respect to the horizontal radiation, with the latter remaining almost equal to 
the static case (no horizontal motions, no horizontal Doppler brightening). This
velocity-induced limb darkening is the origin of the anisotropy enhancement. 
 
However, note that the maximum anisotropy does not rise indefinitely when
increasing the maximum velocity. If the velocity gradient in units of the 
Doppler width is larger than $\sim 3$ (see curves for $\xi_{\rm{max}}=5$ in Fig. \ref{fig:me_fig}), the
anisotropy at the surface saturates and decreases (even below the curves
corresponding to shorter velocity gradients). It is accompanied by a bump around
$\tau_l=1$ when the maximum velocity gradient is taking place at low density
layers ($\tau_0 \lesssim 1$). This behavior can be understood using Eq. (\ref{eq03}). 
When $\xi_{\rm{max}} \lesssim 3$, an increment in
$\xi_{\rm{max}}$ entails a rise in $\bar{J}^0_0$, $\bar{J}^2_0$ and
$\bar{J}^2_0/\bar{J}^0_0$ ($w_{\rm{line}}$) in the upper atmosphere, what means
that the velocity gradients enhance the imbalance between vertical and
horizontal radiation. However, if $\xi_{\rm{max}}$ is above that threshold,
$\bar{J}^0_0$ and $\bar{J}^2_0$ rise, but the ratio $\bar{J}^2_0/\bar{J}^0_0$
saturates and diminishes. The reason is that a large velocity gradient makes the
absorption profiles associated to almost horizontal outgoing rays
($0<\mu<1/\sqrt{3}$) shifted so much that they also capture the background
continuum radiation. Their contributions are negative to the angular integral of
$\bar{J}^2_0$ but positive for $\bar{J}^0_0$. 

Separating the contributions of rays with
angles in the range $1/\sqrt{3}<|\mu|<1$ (that we refer to with the label $+$) and angles
in the range $0<|\mu|<1/\sqrt{3}$ (that we refer to with the label $-$), the
line anisotropy can be written as $w_{\rm{line}}=w^+_{\rm{line}}+w^-_{\rm{line}}=\bar{J}^{2+}_0/\bar{J}^0_0-|\bar{
J}^{2-}_0|/\bar{J}^0_0$. Here,
$\bar{J}^0_0$ and $\bar{J}^{2+}_0$ grow always with $\xi_{\rm{max}}$, but
$|\bar{J}^{2-}_0|$ only grows appreciably when $\xi_{\rm{max}} \gtrsim 3$. Therefore, although 
$w_{\rm{line}}$ increases for all velocity gradients, its enhancement is
smaller for large velocity gradients than for smaller ones. This
effect occurs as well when motions take place deeper ($\tau_0\gtrsim 1$)
but it is less important and the anisotropy bump and saturation are reduced.

For the considered velocity fields (with a negligible gradient in the
upper atmosphere), $\bar{J}^0_0$ and $\bar{J}^2_0$ reach an asymptotic value
in optically thin regions. We have verified that this 
effect does not occur if the velocity gradient is not zero in those layers.
In any case, the presence of a large anisotropy in optically thin heights barely
affects the emergent linear polarization profiles.


\subsection{Two-level model atom in dynamic atmospheres.}\label{sub:tla}
Before going to a more realistic case, a final illustrative example is
considered. In this case, we assume the same parameterization of the velocities
than in the previous example, but now we solve the complete iterative RT problem
with a two-level atom model and a specific temperature stratification. Consequently, 
the source function and the anisotropy are consistently obtained
in a moving atmosphere. The intensity source function is
$S_I=r_{\nu\mu}S^{\rm{line}}_I+(1-r_{\nu\mu})B$ \citep[e.g.,][]{rybicki-hummer}, with $r_{\nu\mu}=\phi'_{\ell
u}(\nu,\mu)/(r_c+\phi'_{\ell u}(\nu,\mu))$ and the expression for the
line source function remains formally equal to that of the static case, being
$S^{\rm{line}}_I=(1-\epsilon)\bar{J}^0_0+\epsilon B$, where $B$ is the imposed
Planck function, $\epsilon$ is the inelastic collisional parameter and
$\bar{J}^0_0$ is calculated with Eq. (\ref{eq02}).

\begin{figure*}[htb]
\epsscale{1.0}
\plotone{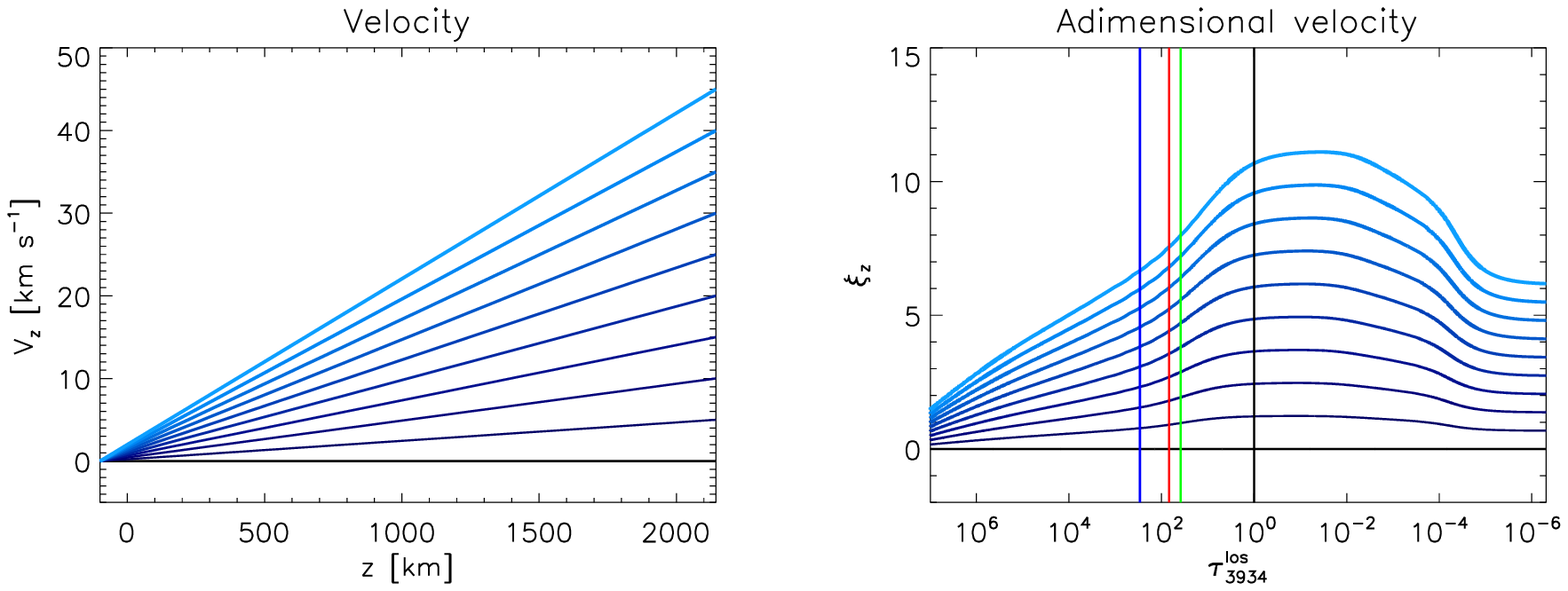}    
\plotone{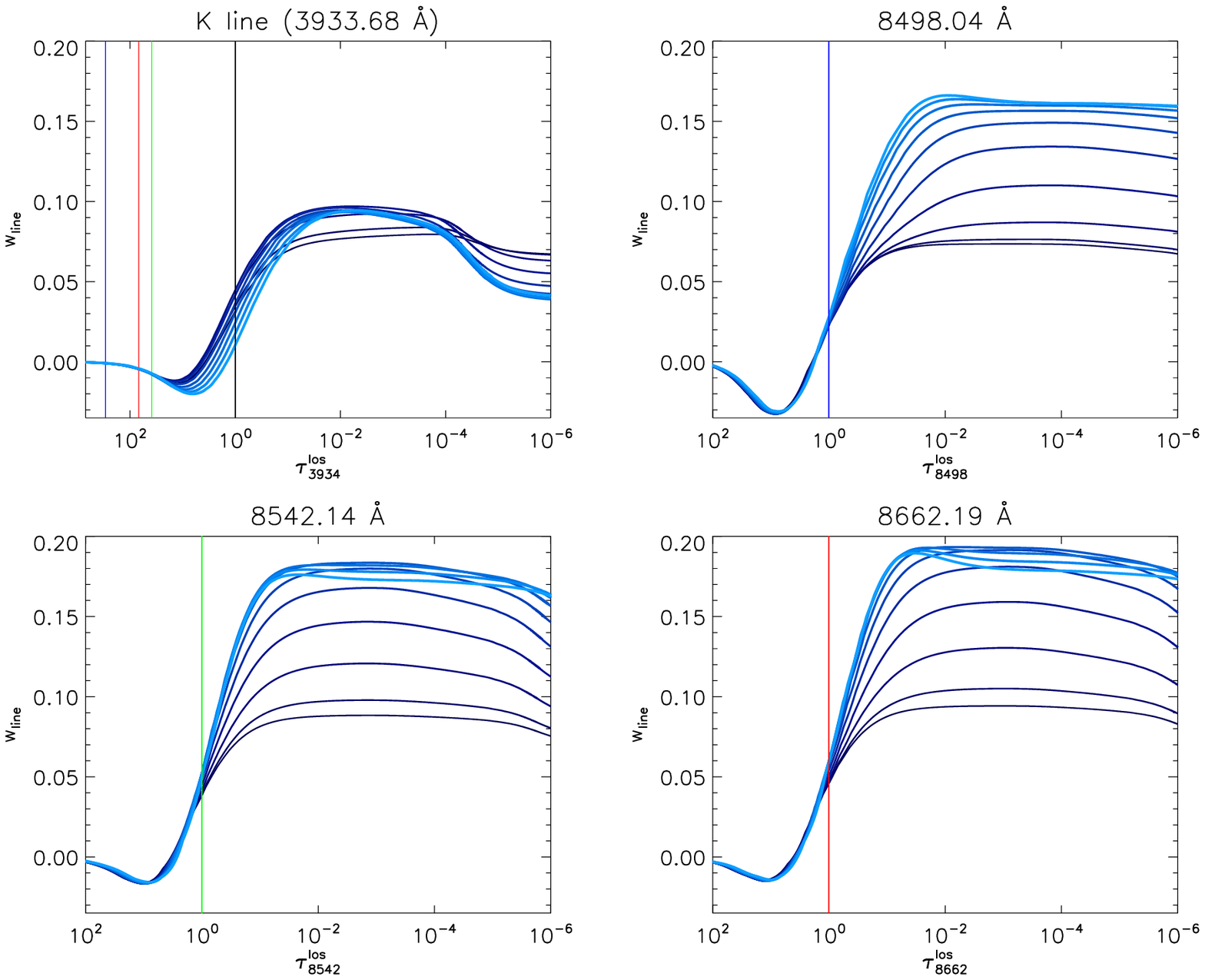}  
\caption{Amplification of the line anisotropy ($\rm{w_{line}}=\sqrt{2} \bar{J}^2_0/\bar{J}^2_0$) due to vertical velocity gradients. 
Upper left panel: linear velocity fields versus height, with velocity 
gradients going from $0$ (darker lines) to $20\,\rm{m \cdot s^{-1} \, km^{-1}}$ 
(light blue lines) in steps of $2.23\,\rm{m \cdot s^{-1} \, km^{-1}}$. Upper right 
panel: corresponding adimensional velocity fields ($\xi_z$) for a FALC 
temperature stratification and a constant microturbulent velocity of 
$3.5 \,\rm{km\, s^{-1}}$. The horizontal axis is in units of the K-line 
optical depth along the line of sight (los). The vertical lines mark the position of 
$\tau^{\rm{los}}_{\nu_0}= 1$ for the transitions $8498 \AA$ (blue), $8542 \AA$ (green), $8662 \AA$ (red) and the K line (black). 
Remaining panels: corresponding line anisotropy factors plotted 
against $\tau^{\rm{los}}_{\nu_0}$ for each line. \label{anisovars}}
\end{figure*}

The qualitative behaviour explained in the previous subsection is maintained in these
two-level atom calculations. For small $\epsilon$ values (large NLTE effects), 
the source function $S_I \thickapprox
\bar{J}^0_0$ shows Doppler brightening effects and its 
surface value depends on the maximum velocity gradient and on the
maximum background continuum set by the photospheric conditions (upper panel in Fig.
\ref{fig:tla1}). The anisotropy rises proportionally to the velocity gradient until
a saturation occurs (lower panel in Fig. \ref{fig:tla1}). A similar behavior is 
found when the maximum velocity gradient occurs higher in the atmosphere (see 
Fig. \ref{fig:tla2} in Appendix \ref{app:C}).

In a static atmosphere, the radiation field anisotropy is dominated by the presence
of gradients in the intensity source function (Trujillo Bueno 2001; Landi Degl'Innocenti \& Landolfi
2004), which can be modified via 
the Planck function (equivalently, the temperature). In the dynamical case that
we are dealing with, the slope of the source function is also modified due to the
existence of velocity gradients thanks to the frequency-decoupling caused
by relative motions between absortion profiles (Doppler brightening). In general,
both mechanisms act together (velocity-induced and temperature-induced modification
of the source function gradient) and the ensuing anisotropy and the emergent linear
polarization profiles are modified accordingly.

It is important to note that the adimensional velocity $\xi$
depends both on the velocity and also on the line Doppler width because
$\xi (\tau)= \delta\nu/\Delta\nu_D= v_z/\sqrt{2k_BT/m}$, with $k_B$ the Boltzmann
constant, $T$ the temperature and $m$ the mass of the atom. In the photosphere,
where velocities are much lower than in the chomosphere, $\xi$ is expected to be
negligible. In the chromosphere, plasma motions are important and the temperature is
still comparable to that of the photosphere, inducing $\xi$ to be controlled by
the velocity field. However, for layers in the transition region and above, the high
temperatures reduce the value of $\xi$. In any case, at these heights, the density is so low that, although $\xi$ (and
consequently the anisotropy) could have a highly variable behaviour, the
emergent polarization profiles of chromospheric lines will not be sensitive to
them.   

\begin{figure*}[!t]
 \epsscale{0.9}
\plotone{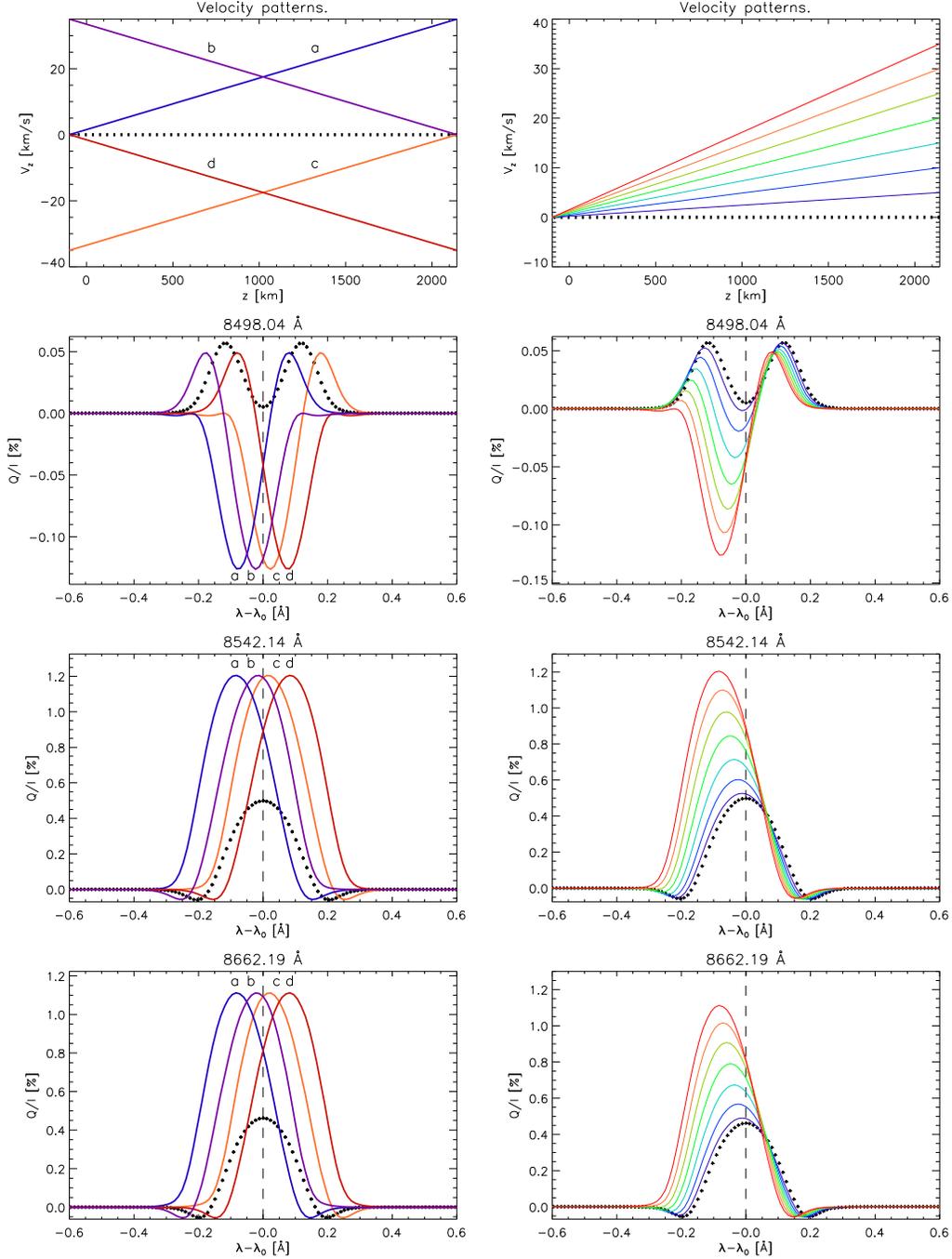}
\caption{Left panels: calculation at $\mu=0.1 $ of the 
emergent $Q/I$ polarization signals of the Ca \textsc{ii} IR triplet when four 
different choices for the vertical velocity gradients are imposed. Positive 
velocities imply upflowing plasma. Gradient ``a''/``b'' simulates an atmosphere 
where the plasma is entirely moving towards the observer increasing/decreasing 
linearly the velocity along the outgoing z axis. Gradients ``c'' and ``d'' are 
the same for plasma moving away from the observer. The black dotted line is the 
solution for the static reference case. Each curve is computed on the converged 
solution of the multilevel NLTE problem described in Sec. \ref{sec:problem}. 
Right panels: same calculations than in the left-hand panels, 
but with different velocity gradient values varying 
from $0$ to $16.3 \,\rm{m \, s^{-1}\,km^{-1}}$ in steps of $2.3 \,\rm{m \, s^{-1}\,km^{-1}}$ 
(see top right panel). These results show that the 
polarization signals are increased and shifted with respect to the static case
depending only on the absolute value of the vertical velocity field gradient and
independently of the sign of the velocity field.
\label{fig:gradsv_ctes}}
\end{figure*}


\section{Results for the Ca {\sc ii} IR Triplet .}\label{sec:anisotriplet}
Now, we study the effect of the velocity field on a multilevel atomic system 
in a realistic atmospheric model, within the more general framework described in Sec. \ref{sec:problem}. 
We consider the formation of the scattering polarization pattern of the Ca {\sc
ii} infrared triplet in a semi-empirical model atmosphere (FAL-C model of
Fontenla, Avrett, and Loeser 1993) in the presence of vertical velocity fields
($\mathbf{v}=v_z(z)\mathbf{k}$, with $\mathbf{k}$ the unit vector along the vertical
pointing upwards). We will assume a constant
microturbulent velocity field of 3.5~km~s$^{-1}$, a representative value for the
region of formation which gives a realistic broadening of the triplet profiles. 
\\

\subsection{Behaviour of the anisotropy in the Ca {\sc ii} IR triplet}\label{subsec:aniso}
For simplicity, we set linear velocity fields (constant velocity gradient
along z) between $z=-100$ and 2150 ~km (see Fig. \ref{anisovars}). Consequently, the adimensional velocity field $\xi_z$ has a non-monotonic behavior 
due to its dependence on the temperature (upper right panel in Fig. \ref{anisovars}). In
the chromosphere, where the Ca \textsc{ii} triplet lines form, $\xi_z$ is dominated by the
macroscopic motions. Here, the velocity gradients produce variations in the anisotropy of the triplet
lines that agree with the behaviour outlined in the previous sections. Namely, an
amplification and a subsequent saturation of the anisotropy factor due to the
significant velocity gradient at those heights (see the lower panels and middle
right panel in Fig. \ref{anisovars}). Above the chromosphere, on the contrary,
the temperature dominates ($\xi_z$ stabilizes and diminishes) and the anisotropy slightly
decreases with height.

If the (adimensional) velocity gradient is negligible
where the line forms (around $\tau^\mathrm{los}_{\nu_0}\thicksim 1$), the anisotropy
remains unaffected. Otherwise, if a spectral line forms at very hot layers,
where the absortion profiles are wider and their sensitivity to the velocity
gradients is lower, the Doppler brightening will not be so efficient
amplifying the anisotropy. This is the case of the anisotropy of the
Ca \textsc{ii} K line (middle left panel in Fig. \ref{anisovars}).  Compare how the slope of $\xi_z$ is smaller where the Ca \textsc{ii} K line forms (black line on Fig. \ref{anisovars}) than where the triplet lines do. Consequently, the enhancement 
of the line anisotropy through the presence of velocity gradients in this line is reduced. 

All our calculations demonstrate that the anisotropy in the Ca {\sc ii}
IR triplet can be amplified through chromospheric vertical velocity gradients.
This results suggest that the same occurs with the ensuing linear polarization
profiles.

\subsection{The impact on the polarization of the emergent radiation.}\label{sec:impact}
For investigating the effect of vertical velocity fields on the emergent fractional
polarization profiles we perform the following numerical experiments. First, we
impose velocity gradients with the same absolute value but opposite signs (top
left panel of Fig. \ref{fig:gradsv_ctes}). The resulting emergent $Q/I$
profiles (remaining left panels of Fig. \ref{fig:gradsv_ctes}) are magnified by a
significant factor ($>2$ for all the transitions) with respect to the static
case (black dotted line). The linear polarization profiles have the same amplitude, independently of the
sign of the velocity gradient. Another remarkable
feature is the asymmetry of the profiles, having a higher blue wing in those
cases in which the velocity gradient is positive and a higher red wing when the
velocity gradient is negative, independently of the velocity sign. Note also
that the $Q/I$ profile is shifted in frequency due to the relative velocity between the
plasma and the observer.

\begin{figure}[htb]
\centering
\epsscale{1.2}
\plotone{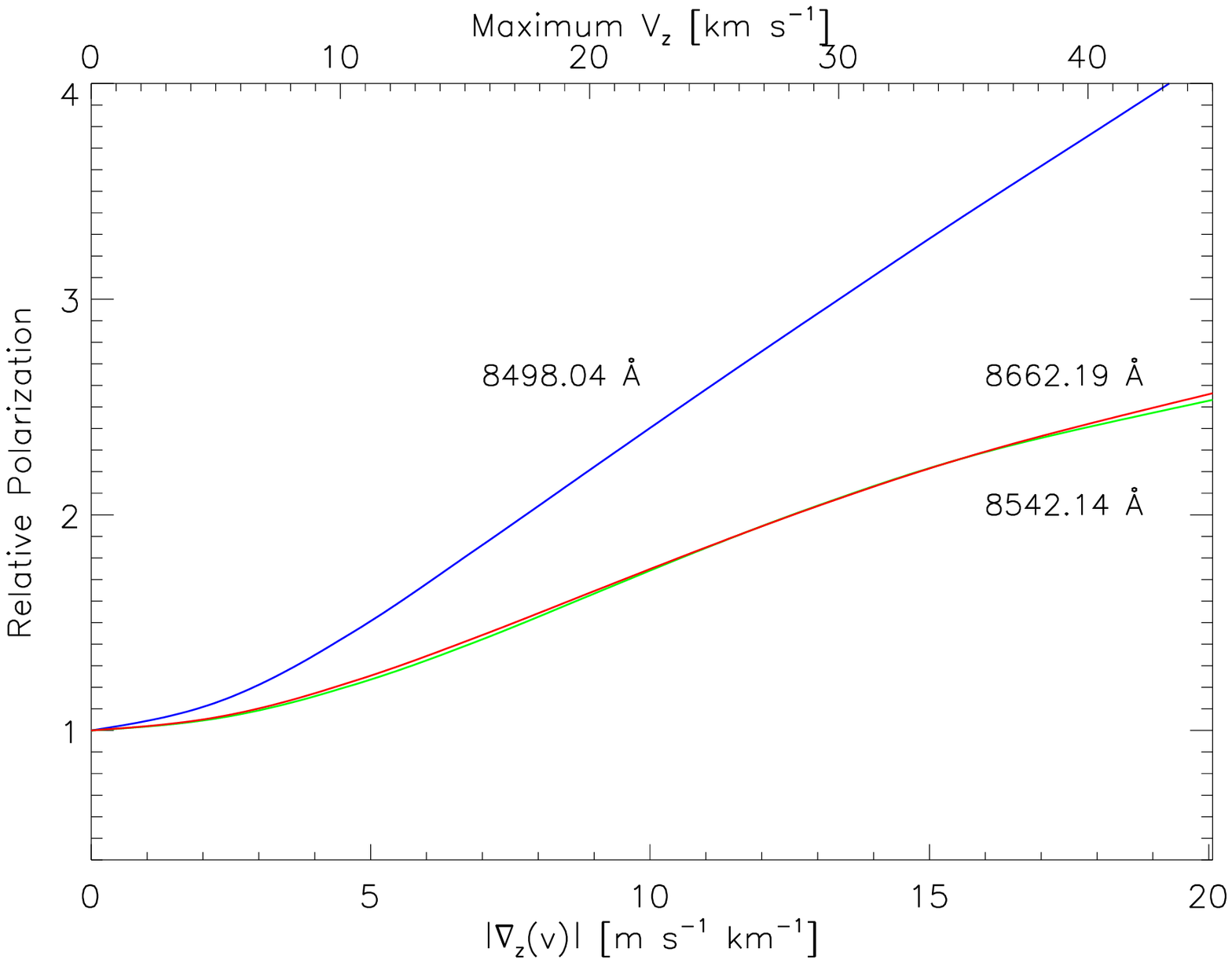}
\caption{$(Q/I)_\mathrm{pp}$ normalized to the static case solution versus the velocity 
gradient (bottom axis) and the maximum absolute velocity (top axis) for linear velocity 
fields appearing in Fig. \ref{fig:gradsv_ctes}. The results are invariant under velocity 
sign changes. The transition $\lambda8498$ is more sensitive to velocity variations due 
to its thinner absortion profile.\label{fig:gradvarlaw}}
\end{figure}

As a second experiment, we consider different velocity fields with increasing
gradients (right upper panel in Fig. \ref{fig:gradsv_ctes}). In the ensuing
$Q/I$ profiles we see that the larger the velocity gradient, the larger 
the frequency shift of the emergent profiles and the larger the amplitude. In
all transitions, one of the lateral lobes of the signal remains almost constant.
Thus, what really changes is the central part of the profiles, being a ``valley''
in the $\lambda8498$ line and a ``peak'' in the other two transitions. To quantify
these variations, we define $(Q/I)_\mathrm{pp}$ (peak-to-peak amplitude of $Q/I$) as the difference between the lowest and the highest value
of the emergent $Q/I$ signal, which is also a measure of its contrast. Note that, as
expected from the first experiment, $(Q/I)_\mathrm{pp}$ depends only on the
absolute value of the gradient. Figure \ref{fig:gradvarlaw} summarizes these
results.

The sensitivity of the linear polarization to the velocity gradient can be measured approximately as
commented in Sec. \ref{sub:scattercloud}, using a parameter
$\alpha=\Delta\nu_D/w$ that accounts for the difference in width of the absorption profile
with respect to the emergent intensity profile. If $\alpha \thicksim 1$, small
adimensional velocities will produce large changes in shape; if
$\alpha \ll 1$, much larger $\xi_z$ values are needed for the same effect. In
the case of the IR triplet lines, $\alpha \thicksim 0.355$ in the formation region
of $\lambda8498$ and around $0.29$ and $0.285$ in the formation region of
$\lambda8542$ and $\lambda8662$ (having wider profiles), respectively. Then, the
former is more sensitive to velocity variations in its formation region (Fig.
\ref{fig:gradvarlaw}). Finally, the K line has $\alpha\,(\tau^{\rm{los}}_{\nu_0}=1)
\thicksim 0.015$, a low value due to its wider spectral wings.  

The enhancement of the polarization signals are a consequence of the increase in the
anisotropy. Therefore, since this increase is produced by the presence of velocity fields, the
polarization signals of the Ca {\sc ii} IR triplet are sensitive also to the dynamic state of
the chromosphere.

\subsection{Variations on the atomic alignment due to velocity
gradients.}\label{sub:EB}
In order to get physical insight on the formation of the emergent polarization
profiles, we use an analytical approximation. Following Trujillo Bueno (2003),
the emergent fractional linear polarization for a strong line at the central wavelength
can be approximated with:
\begin{equation}\label{eq:qestimated2}
\frac{Q}{I} \thickapprox  \frac{3}{2\sqrt{2}} (1-\mu^2) \left[ 
{\sc}w^{(2)}_{J_u J_{\ell}} \cdot \sigma^2_0(J_u)  -  {\sc}w^{(2)}_{J_{\ell} J_u}   \cdot
\sigma^2_0(J_{\ell}) \right].
\end{equation}
The symbols ${\sc}w^{(2)}_{J J'}$ are numerical coefficients already introduced
in Sec. \ref{sec:problem}. The quantities $\sigma^2_0(J_u)$ and
$\sigma^2_0(J_{\ell})$ are the fractional alignment coefficients
($\sigma^2_0=\rho^2_0/\rho^0_0$) evaluated at $\tau^{\rm{los}}_{\nu} = 1$ for the
upper and lower level of the transition, respectively. This is the
generalization of the Eddington-Barbier (EB) aproximation to the scattering
polarization and establishes that changes in linear polarization (for a static
case) are induced by changes in the atomic aligment of the energy levels.

\begin{figure*}[htb]
\epsscale{1.0}
\plotone{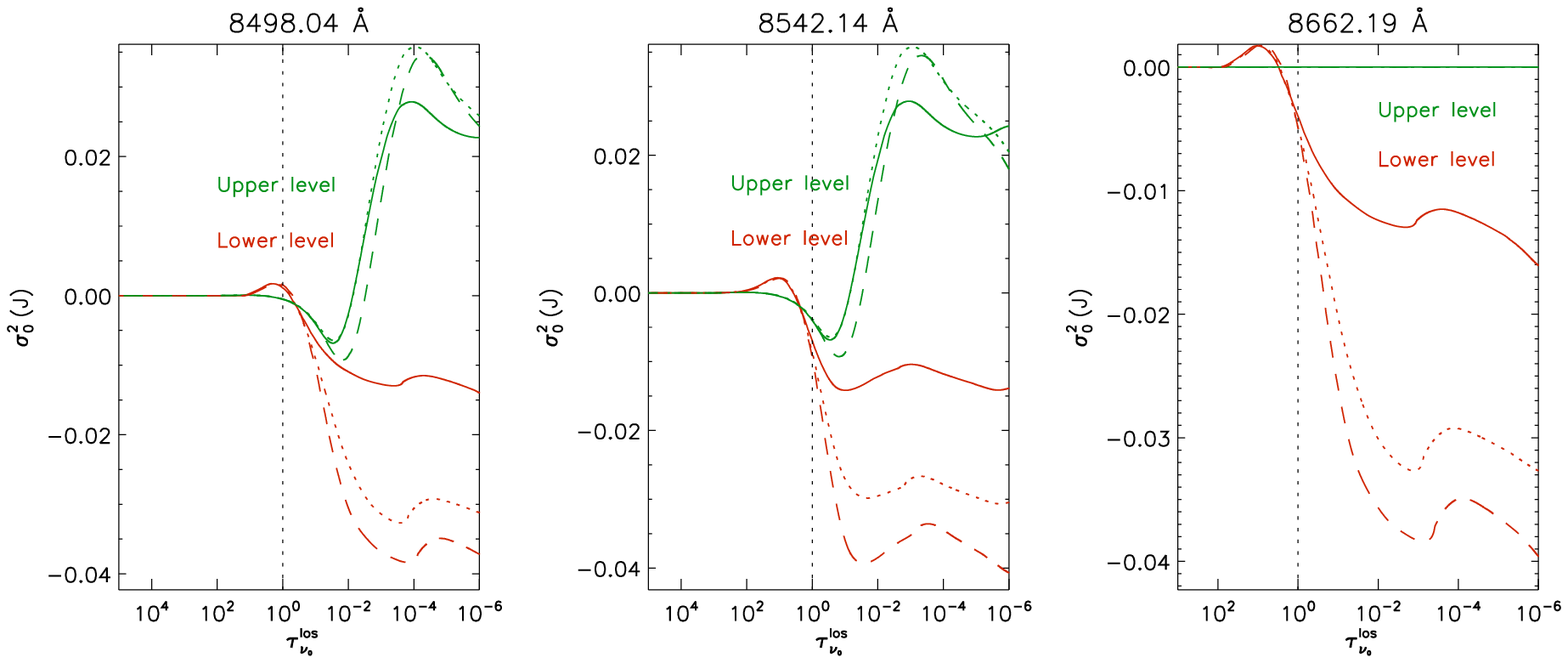}
\caption{Behaviour of the fractional alignments $\sigma^2_0(J_u)$ (green) 
and $\sigma^2_0(J_l)$ (red) of each Ca \textsc{ii} IR triplet transition
for three of the velocity fields shown in Fig. \ref{fig:gradsv_ctes}. The solid lines correspond
to the reference static case. The dotted lines correspond to the case with maximum velocity 
$15\, \rm{km \, s^{-1}}$. The dashed lines are associated with the case with maximum 
velocity $30 \, \rm{km \, s^{-1}}$. The horizontal axis is the line center 
optical depth for each value of the velocity gradient. The vertical dotted 
line marks the height where $\tau^{\rm{los}}_{\nu_0}=1.$\label{fig:rhosqi2}}
\end{figure*}

Our calculations show that vertical velocity fields with moderate gradients
($\lesssim 10\,\rm{m\, s^{-1}  km^{-1}} $ in a linear velocity field, as the ones shown in the figures)
do indeed produce variations in the fractional alignment, which are small for 
$|\sigma^2_0(J_u)|$ and significant for $|\sigma^2_0(J_{\ell})|$ (see Fig. \ref{fig:rhosqi2}). The
lower level alignment is the main driver of the changes produced in the $Q/I$
signals. This is strictly true for the $\lambda8662$ line, whose upper
level with $J=1/2$ cannot be aligned (zero field dichroism polarization).
In the other transitions, a certain influence of the upper level alignment
becomes important only for large gradients. The reason is that the K transition
is so strong in comparison with the IR triplet that it is dictating the common
upper level 5 alignment (Fig. \ref{fig:elevels}). In fact, $\sigma^2_0(J_5)$ is
driven by the K line anisotropy which, at cromospheric heights, is almost
unaffected for the considered velocity gradients, as we discussed in Sec. \ref{subsec:aniso}
(Fig. \ref{anisovars}). Thus, the strong H and K lines feed population to the
upper levels and the K line controls the alignment of the $^2P_{3/2}$ level (see Fig. 1), while the polarization signals
of the IR triplet change with velocity fields affecting $\sigma^2_0(J_{\ell})$ (through the
anisotropy enhancement). 

To illustrate the well-known link between the alignment and the anisotropy we
can follow the next reasoning. For the Ca \textsc{ii} model atom that
we deal with in this work, it is posible to derive a simple
analytic expression that relates the anisotropy and the alignments for the $\lambda8542$
transition. Making use of Eqs. (\ref{see04}) and (\ref{see05}) and neglecting
second order terms and collisions, we find that:
\begin{equation}\label{eq:relacionw}
{2}\sigma^2_0(J_5)-{\sqrt{7}}\sigma^2_0(J_3) \simeq  {\sc w}_{line}(3\rightarrow
5).
\end{equation}
As before, we can roughly assume that $\sigma^2_0(J_5)\thicksim \rm{constant}$
at the chromosphere because it is controled by the K line. Then, Eq.
(\ref{eq:relacionw}) suggests that, if the radiation anisotropy increases at
that heights, an amplification of $|\sigma^2_0(J_3)|$ occurs (note that $\sigma^2_0(J_3)$
is negative for these lines). A more aligned atomic
population produce a more intense scattering polarization signal.

\section{Conclusions}

When vertical velocity gradients exist, the
polarization profiles are always shifted in wavelength, asymmetrized and enhanced in
amplitude with respect to the constant velocity case. The reason is that increments in the absolute value
of the velocity gradient increase the source function (Doppler brightening) and
enhance the anisotropy of the radiation field (Secs. \ref{sec:preview} and
\ref{sec:anisotriplet}), that in turn modify the fractional alignment (Sec.
\ref{sub:EB}) and amplify the scattering polarization profiles (Sec.
\ref{sec:impact}).
 
For this very reason, all calculations assuming static models in the
formation region might underestimate the scattering polarization amplitudes and
not capture the right shape of the profiles. In particular, it must be taken into
account that the Ca {\sc ii} IR triplet lines form under non-LTE conditions in
chromospheric regions, where velocity gradient may be significantly large due to the upward
propagation of waves in a vertically stratified atmosphere (e.g., Carlsson \& Stein 1997). Probably, in
photospheric and transition region lines the effect of velocities on polarization can be
safely neglected (they will be predominantly amplified by temperature gradients
as discussed in Sec. \ref{sub:tla}), but not necessarily in the chromosphere. In our study
we see that the $\lambda8498$ line is more sensitive to macroscopic motions in the 
low-chromosphere, while the $\lambda8542$ and $\lambda8662$ lines are especially amplified when
strong velocity gradients are found at heights around $1.5 \, \rm{Mm}$ and higher in our
model. 

At the light of these results, it is obvious that the effect of the velocity might be
of relevance for measuring chromospheric magnetic fields. In
particular, the described mechanism might turn out to be important for the correct
interpretation of polarization signals in the Sun with the Hanle effect. Given
that weak chromospheric magnetic fields are inferred with the Hanle effect using the difference
between the observed linear polarization signal and the signal
that would be produced in the absence of a magnetic field, it is crucial
to correctly compute the reference no-magnetic signal. In the
Ca {\sc ii} IR triplet, it could be possible to break the degeneracy of the
combined effects by taking into account the different sensitivities of the three
lines of the triplet to the magnetic field and the thermodynamics. As stated by MSTB2010,
the $\lambda8498$ line is very sensitive to the thermal structure of the
atmosphere. Likewise, the response of this line to the velocity
gradient is also higher than in the other two. Then, a first step could be to
characterize the response of all the lines to combined variations of the
temperature, velocity and magnetic field and find observables
(i.e., amplitude ratios) that are as insensitive to the temperature and
velocity as possible and as sensitive to the magnetic field as possible. We 
are currently carrying out this study on realistic velocity fields,
including shocks and temporal variations.

The polarization amplification mechanism that we have discussed in this 
paper is not limited to plane-parallel atmospheres, although its
effect is surely more important in plane-parallel atmospheres than
in three-dimensional ones. The reason is that i) gradients in a
three-dimensional atmosphere are weaker given the
increased degrees of freedom and ii) non-resolved
motions or large variations in velocity direction along the medium mix the
contribution of different layers and broaden the profiles, diminishing them in
amplitude. In any case, strong three-dimensional velocity gradients
might create preferred directions along which the plasma becomes more optically
thin through the radiative uncoupling mechanism discussed in this paper.


The amplification of the radiation field anisotropy through vertical velocity
gradients is a general and interesting phenomenon that improves our
understanding of the stellar atmospheres. With the present investigation we have
obtained some feeling about the formation of the Ca {\sc ii} IR triplet in dynamic situations.

\appendix 

\section{Special functions in section 3}\label{app:A}

Introducing Equation~\ref{eq04} into Equation~\ref{eq02}
\begin{equation}\label{eqa01}
\bar{J}^0_0=\frac{1}{2}\int_0^\infty d\nu'
\frac{1}{\sqrt{\pi}\Delta\nu_D}\exp\{-(\frac{\nu'-\nu_0}{\Delta\nu_D})^2\}
\int_0^1d\mu
I^{(0)}(1-u+u\mu)[1-a\exp\{-(\frac{\nu'(1+v_z\mu/c)-\nu_0}{w})^2\}].
\end{equation}
Introducing the variables $x=(\nu'-\nu_0)/w$, $\alpha=\Delta\nu_D/w$, and
$\xi=v_z\nu_0/(cw)$, then the mean intensity in the comoving frame
\begin{equation}\label{eqa02}
\frac{\bar{J}^0_0}{I^{(0)}}=\frac{1}{2}\int_{-\infty}^{\infty}dx
\frac{1}{\sqrt{\pi}\alpha}{\rm e}^{-(x/\alpha)^2}
\int_0^1d\mu (1-u+u\mu)(1-a\exp^{-(x+\xi\mu)^2}).
\end{equation}
In passing from Equation~(\ref{eqa01}) to Equation~(\ref{eqa02}), we have
extended the integration limit on $x$ to $\infty$. 
Analogously for the anisotropy in the comoving frame
\begin{equation}\label{eqa03}
\frac{\bar{J}^2_0}{I^{(0)}}=\frac{1}{4\sqrt{2}}\int_{-\infty}^{\infty}dx
\frac{1}{\sqrt{\pi}\alpha}{\rm e}^{-(x/\alpha)^2}
\int_0^1d\mu (3\mu^2-1)(1-u+u\mu)(1-a{\rm e}^{-(x+\xi\mu)^2}).
\end{equation}

The following integrals are easily evaluated (see Spiegel \& Liu, 1998)
\begin{align}\label{eqa04}
{\cal I}_0(\alpha;\xi)&\equiv\int_{-\infty}^{\infty}dx
\frac{1}{\sqrt{\pi}\alpha}{\rm e}^{-(x/\alpha)^2}
\int_0^1 (1-u+u\mu)[1-a {\rm e}^{-(x+\xi\mu)^2}]d\mu \notag \\
&=\frac{1}{2}[2-u+a(u-1)\sqrt{\pi}\frac{1}{\xi}{\rm
Erf}(\frac{\xi}{\sqrt{1+\alpha^2}})
+a u\sqrt{1+\alpha^2}\frac{1}{\xi^2}(\exp\{-\frac{\xi^2}{1+\alpha^2}\}-1)], 
\end{align}
\begin{align}\label{eqa05}
{\cal I}_2(\alpha;\xi)&\equiv\int_{-\infty}^{\infty}dx
\frac{1}{\sqrt{\pi}\alpha}{\rm e}^{-(x/\alpha)^2}
\int_0^1 (3\mu^2-1)(1-u-u\mu)[1-a {\rm e}^{-(x+\xi\mu)^2}]d\mu \notag \\
&=\frac{1}{4}[u-2 a (u-1) \sqrt{\pi} \frac{1}{\xi} {\rm
Erf}(\frac{\xi}{\sqrt{1+\alpha^2}}) 
+2a\sqrt{1+\alpha^2}\frac{1}{\xi^2}
\left[u+(3-u)\exp\{-\frac{\xi^2}{1+\alpha^2}\}\right] \\
&+3a(u-1)(1+\alpha^2)\sqrt{\pi}\frac{1}{\xi^3}{\rm
Erf}(\frac{\xi}{\sqrt{1+\alpha^2}})
+6au(1+\alpha^2)^{3/2}\frac{1}{\xi^4}(\exp\{-\frac{\xi^2}{1+\alpha^2}\}-1)],
\notag
\end{align}
where we have made use of 
\begin{equation}
\int_{-\infty}^{\infty}dx \frac{1}{\sqrt{\pi}\alpha}{\rm e}^{-(x/\alpha)^2}
[1-a{\rm e}^{-(x+\xi\mu)^2}]=
1-\frac{a}{\sqrt{1+\alpha^2}}\exp\{-\frac{\mu^2\xi^2}{1+\alpha^2}\}.
\end{equation}
From them, the values for $\bar{J}^0_0$ and the anisotropy
$\sqrt{2}\bar{J}^2_0/\bar{J}^0_0$ are trivially derived.

In the high velocity limit ($\xi\rightarrow\infty$), ${\cal
I}_0(\alpha;\xi)=(2-u)/2$, and ${\cal I}_2(\alpha;\xi)=u/4$ (regardless of
$\alpha$).
In the low velocity limit:
\begin{align}
{\cal
I}_0(\alpha;\xi)&=\frac{1}{2}(1-\frac{a}{\sqrt{1+\alpha^2}})(2-u)+\frac{a(4-u)}{
12(1+\alpha^2)^{3/2}}\xi^2 + {\rm O}(\xi^3), \\
{\cal I}_2(\alpha;\xi)&=\frac{u}{4}(1-\frac{a}{\sqrt{1+\alpha^2}})+
\frac{a(16-u)}{60(1+\alpha^2)^{3/2}}\xi^2 + {\rm O}(\xi^3).
\end{align}

\section{Statistical equilibrium equations. }\label{app:B}
The rate equations for the considered problem are as follow. They have been obtained by particularizing to the model-atom of Fig.1 the equations contained in Sects. 7.2 and 7.13 of {Landi Degl'Innocenti} \& {Landolfi} (2004).
\begin{align}
\begin{split}\label{see01}  
\frac{\rm d}{{\rm d}t} \rho^0_0(1) =& -\biggl[\sum_{u=4}^5
B_{1u}\bar{J}^0_0(1\rightarrow u) + \sum_{i\neq 1} C_{1i} \biggr] \rho^0_0(1)
 + A_{41}\rho^0_0(4) + \sqrt{2} A_{51}\rho^0_0(5) +
\sum_{i\neq 1} C_{i1} \sqrt{\frac{2J_i+1}{2}}  \rho^0_0(i),
\end{split}
\displaybreak[0] \\
\begin{split}\label{see02}
\frac{\rm d}{{\rm d}t} \rho^0_0(2) =& -\biggl[\sum_{u=4}^5
B_{2u}\bar{J}^0_0(2\rightarrow u) + \sum_{i\neq 2} C_{2i} \biggr] \rho^0_0(2) 
- \biggl( \frac{1}{\sqrt{2}} B_{24}\bar{J}^2_{0}(2\rightarrow 4)
-\frac{2\sqrt{2}}{5} B_{25}\bar{J}^2_{0}(2\rightarrow 5) \biggr) \rho^2_{0}(2)
\\
& + \frac{1}{\sqrt{2}} A_{42}\rho^0_0(4) + A_{52}\rho^0_0(5) + \sum_{i\neq 2}
C_{i2} \frac{\sqrt{2J_i+1}}{2}  \rho^0_0(i),
\end{split}
\displaybreak[0] \\
\begin{split}\label{see03}
\frac{\rm d}{{\rm d}t} \rho^0_0(3) =& -\biggl[B_{35}\bar{J}^0_0(3\rightarrow 5)
+ \sum_{i\neq 3} C_{3i} \biggr] \rho^0_0(3) 
- \frac{\sqrt{7}}{5} B_{35}\bar{J}^2_0(3\rightarrow 5) \rho^2_0(3) +
\sqrt{\frac{2}{3}} A_{53} \rho^0_0(5) + \sum_{i\neq 3} C_{i3}
\sqrt{\frac{2J_i+1}{6}}  \rho^0_0(i),
\end{split}
\displaybreak[0] \\
\begin{split}\label{see04}
\frac{\rm d}{{\rm d}t} \rho^0_0(4) =& -\biggl[ \sum_{l=1}^2 A_{4l} + \sum_{i\neq
4} C_{4i} \biggr] \rho^0_0(4) 
+ \sum_{l=1}^2 B_{l 4} \bar{J}^0_0(l \rightarrow 4) \sqrt{\frac{2J_l+1}{2}}
\rho^0_0(l) \\
&  + B_{24}  \bar{J}^2_0(l \rightarrow 4) \rho^2_0(2) + \sum_{i\neq 4} C_{i4}
\sqrt{\frac{2J_i+1}{2}}  \rho^0_0(i),
\end{split}
\displaybreak[0] \\
\begin{split}\label{see05}
\frac{\rm d}{{\rm d}t} \rho^0_0(5) =& -\biggl[ \sum_{l=1}^3 A_{5l} + \sum_{i\neq
5} C_{5i} \biggr] \rho^0_0(5)
+ \sum_{l=1}^3 B_{l 5} \bar{J}^0_0(l \rightarrow 5) \frac{\sqrt{2J_l+1}}{2}
\rho^0_0(l) \\
&  -\frac{2\sqrt{2}}{5} B_{25} \bar{J}^2_0(2\rightarrow 5) \rho^2_0(2)
   +\frac{\sqrt{42}}{10}  B_{35} \bar{J}^2_0(3\rightarrow 5) \rho^2_0(3) 
+ \sum_{i\neq 5} C_{i5} \frac{\sqrt{2J_i+1}}{2}  \rho^0_0(i),
\end{split}
\displaybreak[0] \\
\begin{split}\label{see06}
\frac{\rm d}{{\rm d}t} \rho^2_0(2) =& -\biggl[ \sum_{u=4}^5
B_{2u}\bar{J}^0_0(2\rightarrow u) + \sum_{i\neq 2} C_{2i} +D_2^{(2)}\biggr]
\rho^2_0(2) -  \biggl( \frac{1}{\sqrt{2}} B_{24}\bar{J}^2_{0}(2\rightarrow 4)
-\frac{2\sqrt{2}}{5} B_{25}\bar{J}^2_{0}(2\rightarrow 5) \biggr) \rho^0_0(2) \\
& + \frac{1}{5} A_{52}\rho^2_0(5) + \sum_{i=3, 5} C_{i2}^{(2)}
\frac{\sqrt{2J_i+1}}{2}  \rho^2_0(i),
\end{split}
\displaybreak[0] \\
\begin{split}\label{see07}
\frac{\rm d}{{\rm d}t} \rho^2_0(3) =& -\biggl[B_{35}\bar{J}^0_0(3\rightarrow 5)
+ \sum_{i\neq 3} C_{3i} +D_3^{(2)}\biggr] \rho^2_0(3) 
- B_{35} \bar{J}^2_{0}(3\rightarrow 5)\frac{\sqrt{7}}{5} \rho^0_0(3) \\
& + B_{35}\bar{J}^2_{0}(3\rightarrow 5) \biggl[
\sqrt{\frac{5}{7}}\rho^2_{0}(3)
- \frac{9}{2}\sqrt{\frac{3}{35}} \rho^4_{0}(3) \biggr]  +
\frac{2}{5}\sqrt{\frac{7}{3}} A_{53} \rho^2_0(5) + \sum_{i=2, 5} C_{i3}^{(2)}
\sqrt{\frac{2J_i+1}{6}}  \rho^2_0(i),
\end{split}
\displaybreak[0] \\
\begin{split}\label{see08}
\frac{\rm d}{{\rm d}t} \rho^2_0(5) =& -\biggl[\sum_{l=1}^3 A_{5l} + \sum_{i\neq
5} C_{5i} +D_5^{(2)}\biggr] \rho^2_0(5) 
 + \frac{1}{5}B_{25}\bar{J}^0_0(2\rightarrow 5)\rho^2_0(2) +
\frac{\sqrt{21}}{5}B_{35}\bar{J}^0_0(3\rightarrow 5)\rho^2_0(3) \\
& +\frac{1}{2}B_{15}\bar{J}^2_0(1\rightarrow 5)\rho^0_0(1) - \frac{2\sqrt{2}}{5}
B_{25} \bar{J}^2_{0}(2\rightarrow 5) \rho^0_0(2)
                        +\frac{\sqrt{3}}{10} B_{35} \bar{J}^2_{0}(3\rightarrow
5) \rho^0_0(3) \\
& + 2\sqrt{\frac{7}{5}} B_{25} \bar{J}^2_{0}(2\rightarrow 5) \rho^2_{0}(2) -
\sqrt{\frac{3}{5}} B_{35} \bar{J}^2_{0}(3\rightarrow 5) \rho^2_{0}(3)  \\
& +\frac{9}{\sqrt{5}} B_{35} \bar{J}^2_{0}(3\rightarrow 5) \rho^4_{0}(3)+
\sum_{i=2, 3} C_{i5}^{(2)} \frac{\sqrt{2J_i+1}}{2}  \rho^2_0(i),
\end{split}
\displaybreak[0] \\
\begin{split}\label{see09}
\frac{\rm d}{{\rm d}t} \rho^4_0(3) =& -\biggl[ B_{35}\bar{J}^0_0(3\rightarrow 5)
+ \sum_{i\neq 3} C_{3i} +D_3^{(4)}\biggr] \rho^4_0(3) - B_{35}
\bar{J}^2_0(3\rightarrow 5) \biggl[ \frac{9}{2}\sqrt{\frac{3}{35}}
\rho^2_{0}(3) 
+3\sqrt{\frac{11}{70}} \rho^4_{0}(3) \biggr],
\end{split}
\end{align}
where $A_{u{\ell}}$ and $B_{{\ell}u}$ are the Einstein emission and absortion
coefficients; $C_{{\ell} u}$ and $C_{u{\ell}}$ are the excitation and deexcitation
inelastic collisional rates, respectively; $C^{(2)}_{{\ell} u}$ and $C^{(2)}_{u{\ell}}$
are the collisional transfer rates for alignment between polarizable levels
(with $J>1/2$); and $D^{(K)}_{i}$ is the depolarization rate of the $K$-th
multipole of level ${i}$ due to elastic collisions with neutral hydrogen atoms.
The $\rho^K_{0}$ elements are referred to a coordinate system with the
quantization axis along the solar local vertical direction.

\section{Two-level atom calculation in a moving atmosphere. }\label{app:C}

Figure C1 is similar to Fig. \ref{fig:tla1}, but for
velocity fields with $\xi=5$ and maximum gradients occuring at different positions 
along the atmosphere.
\begin{figure}[htb]
\epsscale{0.8}
\plotone{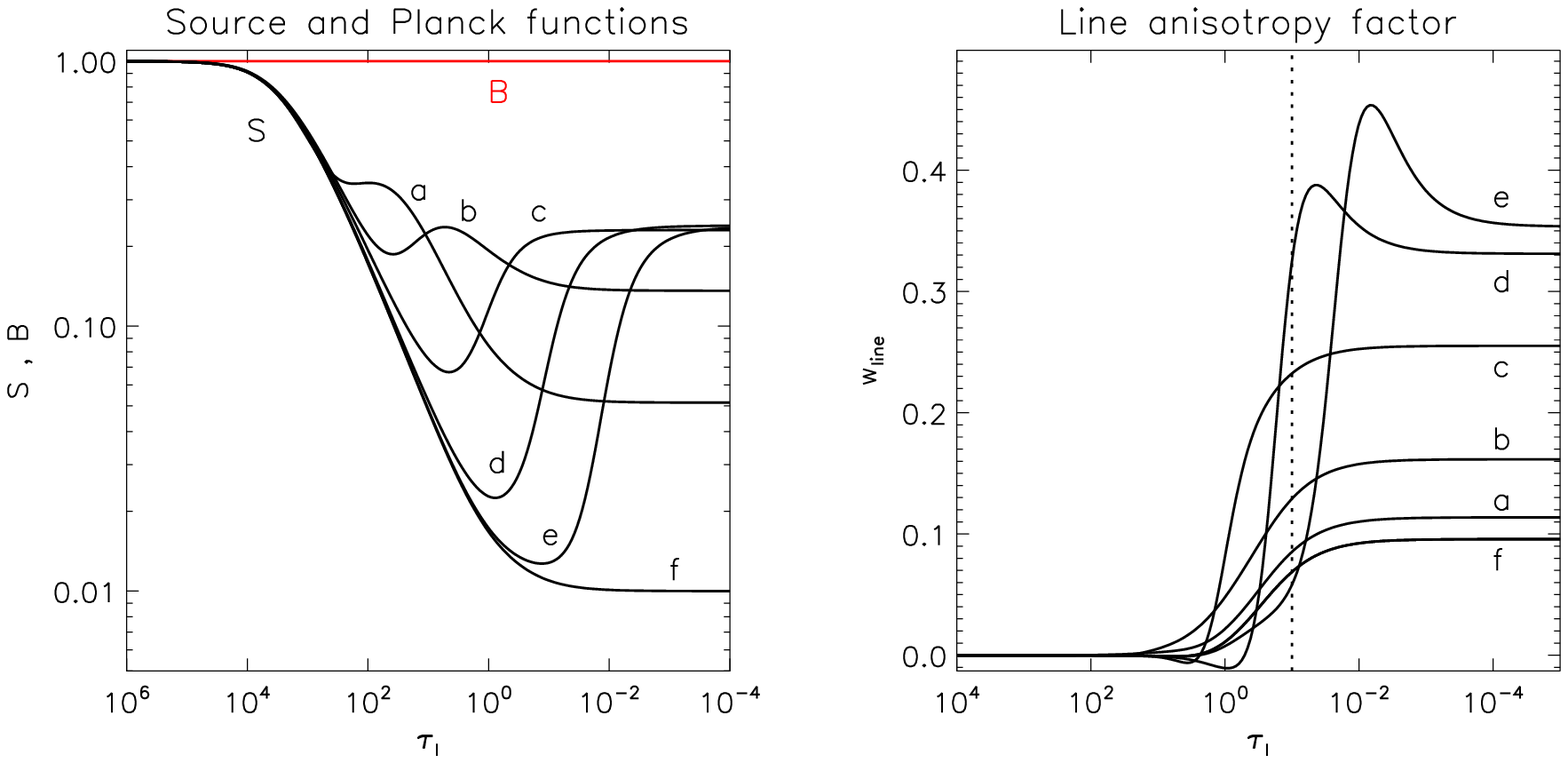}
\plotone{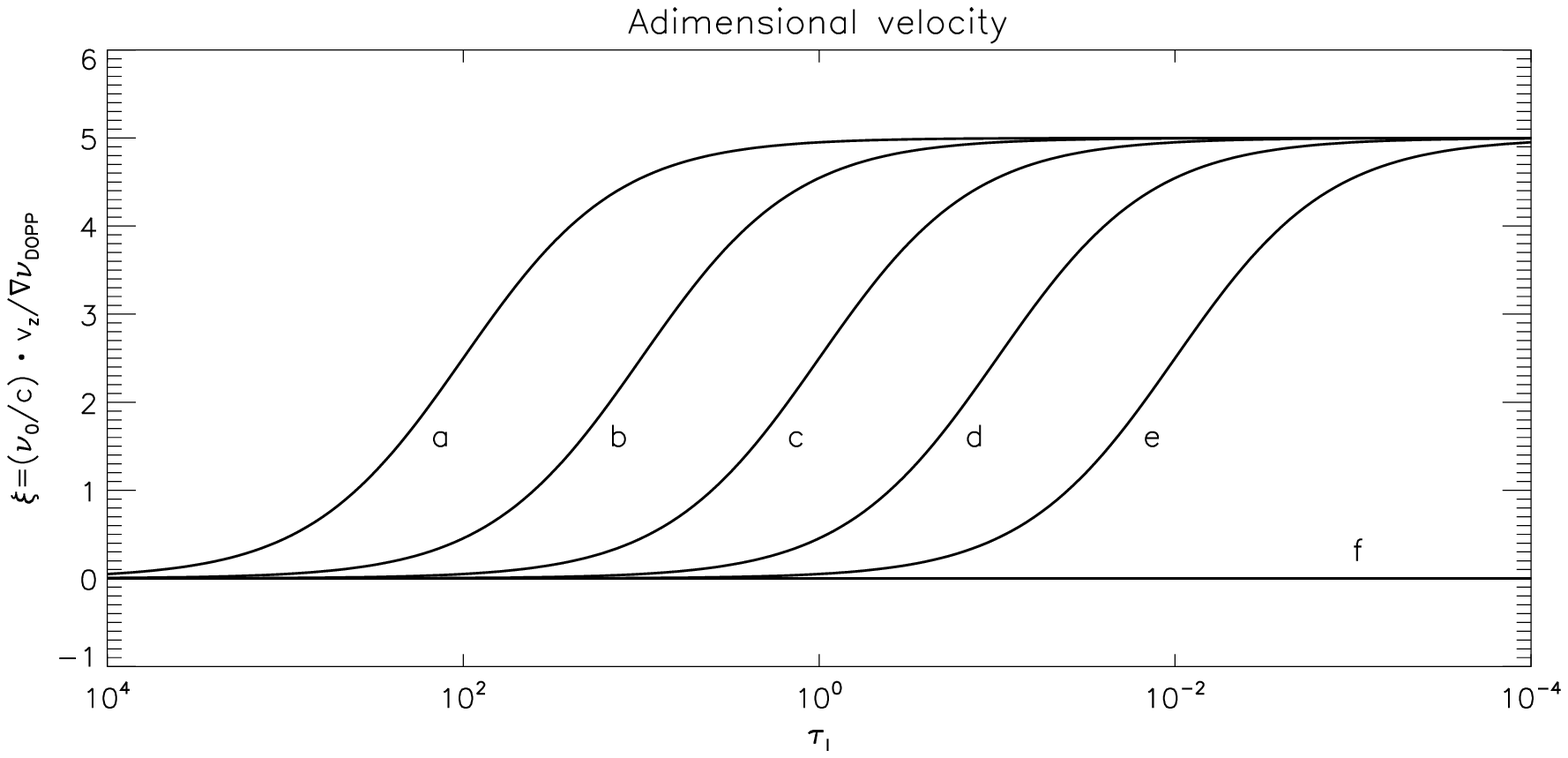}
\caption{Calculations in isothermal two-level atom moving atmospheres 
with $\xi_{max}=5$ . We assume a very strong line ($r_c=0$)  and $\epsilon=10^{-4}$. 
The highest velocity gradient occur at $\tau_l=\tau_0=100,10,1,0.1,0.01$ for a, b, c, d and e, respectively.
The case f corresponds to the solution in a static atmosphere. The vertical dotted line marks the position of  $\tau_l=1$.\label{fig:tla2}
}
\end{figure}

\newpage 

\acknowledgments Financial support by the Spanish Ministry of Science and Innovation through projects AYA2010-18029 (Solar Magnetism and Astrophysical Spectropolarimetry) and CONSOLIDER INGENIO CSD2009-00038 (Molecular Astrophysics: The Herschel and Alma Era) is gratefully acknowledged.


\bibliographystyle{apj} \bibliography{apjmnemonic,../biblio}

\begin{thebibliography}{12}
\expandafter\ifx\csname natexlab\endcsname\relax\def\natexlab#1{#1}\fi
	

\bibitem[{{Brink} \& {Satchler}(1968)}]{angular} 
{Brink}, D.~M., \& {Satchler}, G.~R. 1968, {Angular Momentum}, 2nd edition, Clarendon Press, Oxford

  
\bibitem[{{Carlsson \& Stein}(1997)}]{CS97}
Carlsson, M. \& Stein, Robert F.,	1997 ApJ, 481,500C
	

\bibitem[{{Fontenla} {et~al.}(1993){Fontenla}, {Avrett}, \& {Loeser}}]{fontenla93}
{Fontenla}, J.~M., {Avrett}, E.~H., \& {Loeser}, R. 1993, \apj, 406, 319

\bibitem[{ Spiegel, M. \& Liu, J.}(1998)]{tablas1998}
{Spiegel},M. \& {Liu}, J.,1998 , Mathematical handbook of formulas and tables, Ed. McGraw-Hill.


\bibitem[{{Kunasz} \& {Auer}(1988)}]{kunasz-auer}
{Kunasz}, P., \& {Auer}, L.~H. 1988, JQSRT, 39, 67


\bibitem[{{Landi Degl'Innocenti} (1984)}]{landi84}
{Landi Degl'Innocenti}, E., 1984, Solar Physics, 91, 1

\bibitem[{{Landi Degl'Innocenti} \& {Landolfi}(2004)}]{ll04}
{Landi Degl'Innocenti}, E., \& {Landolfi}, M. 2004, {Polarization in Spectral
  Lines} (Dordrecht: Kluwer)

\bibitem[{{Manso Sainz} \& {Trujillo Bueno}(2003)}]{manso-trujillo-03}
{Manso Sainz}, R., \& {Trujillo Bueno}, J. 2003, Phys. Rev. Letters, 91, 111102

\bibitem[{{Manso Sainz} \& {Trujillo Bueno}(2010)}]{MSTB2010}
{Manso Sainz}, R. \& {Trujillo Bueno}, J. 2010 ApJ...722.1416M

\bibitem[{Mihalas(1978)}]{Mih78} 
Mihalas, D. 1978, Stellar Atmospheres, 2nd Ed. (San Francisco: W. H.
Freeman \& Comp.)

\bibitem[{{Rybicki} \& {Hummer}(1992)}]{rybicki-hummer}
{Rybicki}, G.~B. \& {Hummer}, D.~G., 1992, A\&A, 262, 209

	
\bibitem[{{Stenflo}(1998)}]{stenflo98}
{Stenflo}, J.O. 1998,  A\&A , 338, 301S


\bibitem[{{Stepan} \& {TrujilloBueno}(2010)}]{stepan-jtb}
{\v{S}t\v{e}p\'{a}n}, J. \& {Trujillo Bueno}, J., 2010, Mem. S.A.It. Vol. 81, 810

	
\bibitem[{{Trujillo Bueno}(2001)}]{jtb-sacpeak}
{Trujillo Bueno}, J. 2001, in Advanced Solar Polarimetry,
ed. {M.~Sigwarth}, ASP Conf. Series Vol. 236, 161

\bibitem[{{Trujillo Bueno}(2010)}]{review2010}
{Trujillo Bueno}, J. 2010, in Magnetic Coupling between the Interior and Atmosphere of the Sun,
eds. S.~S.~Hasan \& R.~J.~Rutten, 118T


\bibitem[{{Uitenbroek} \& {Socas-Navarro}(2004)}]{uit-socas-04}
{Uitenbroek}, H. \& {Socas-Navarro}, H. 2004, ApJ, 603, 129

\bibitem[{{Uitenbroek}(2011)}]{uiten11} Uitenbroek, H. 2011, in Solar Polarization Workshop 6, ASP Conf. Ser. Vol. 437, 439



%
%


%
%
%
%
%
  
%
%
%
%
%
%
%
%
%
%
%
%
 
%
%
%

%
%
%
%
%
%
%
%
%
\end{thebibliography}

\end{document}